\documentclass[journal]{IEEEtran}
\usepackage[font=footnotesize,labelfont=bf]{caption}
\ifCLASSINFOpdf
\else
\fi
\usepackage{mdframed}
\usepackage{graphicx}
\usepackage{subcaption}
\usepackage{multirow}
\usepackage{multicol}
\usepackage{listings}
\usepackage{xcolor}
\usepackage{tcolorbox}
\usepackage{caption}
\usepackage{amsmath}
\usepackage{amsfonts}
\usepackage{bbold}
\usepackage{amssymb}
\usepackage{algorithm}
\usepackage{algpseudocode}
\usepackage{booktabs}
\usepackage{algpseudocode}
\usepackage{hyperref}

\definecolor{cgreen}{RGB}{0, 176, 80}

\makeatletter
\def\amsbb{\use@mathgroup \M@U \symAMSb}
\makeatother

\tcbuselibrary{breakable,skins,listings}

\lstdefinestyle{mystyle}{
    backgroundcolor=\color{gray!10},
    commentstyle=\color{green!50!black},
    keywordstyle=\color{blue},
    stringstyle=\color{red},
    basicstyle=\ttfamily\small,
    breaklines=true,
    breakatwhitespace=false,
    showstringspaces=false,
    numbers=left,
    numberstyle=\tiny\color{gray},
    numbersep=5pt,
}

\begin{document}

%\title{AdaCoder: Adaptive Planning Framework for Multi-Agent Function-Level Code Generation} 
\title{AdaCoder: An Adaptive Planning and Multi-Agent Framework for Function-Level Code Generation} 
%\title{AdaCoder: A Multi-Agent Framework for Function-Level Code Generation with Adaptive Planning} 

\author{Yueheng Zhu, Chao Liu, Xuan He, Xiaoxue Ren, Zhongxin Liu, Ruwei Pan, Hongyu Zhang% <-this % stops a space

\thanks{Yueheng Zhu, Chao Liu, Xuan He, Ruwei Pan, Hongyu Zhang were with Chongqing University, Chongqing,
China (e-mail: zhuyueheng@stu.cqu.edu.cn, liu.chao@cqu.edu.cn, xuanhe@stu.cqu.edu.cn,  panruwei@stu.edu.cqu.cn, hyzhang@cqu.edu.cn). Xiaoxue Ren and Zhongxin Liu were with Zhejiang University, Hangzhou, China (e-mail: xxren@zju.edu.cn, liu\_zx@zju.edu.cn)
}

\thanks{Chao Liu is the corresponding author.}
% <-this % stops a space
% \thanks{Xiaoxue Ren was with Zhejiang University, Hangzhou, China (e-mail: xxren@zju.edu.cn)}% <-this % stops a space
\thanks{Manuscript received xx xx, 2025; revised xx xx, 2025.}}


\markboth{Arxiv Preprint}{Zhu \MakeLowercase{\textit{et al.}}: Bare Demo of IEEEtran.cls for IEEE Journals}



\maketitle
%Function-level code generation leverages Large Language Models (LLMs) to automatically produce source code to improve software development productivity. %However, for complex programming tasks,  LLMs often show unsatisfactory performance. 
Recently, researchers have proposed many multi-agent frameworks for function-level code generation, which aim to
improve software development productivity by automatically generating function-level source code based on task descriptions. 
A typical multi-agent framework
consists of Large Language Model (LLM)-based agents that are responsible for task planning, code generation, testing, debugging, etc. 
Studies have shown that existing multi-agent code generation frameworks perform well on ChatGPT. However, their generalizability across other foundation LLMs remains unexplored systematically.
In this paper, we report an empirical study on the generalizability of four state-of-the-art multi-agent code generation frameworks across six open-source LLMs with varying parameter sizes, architectures, and performance levels. Our study reveals the unstable generalizability of existing frameworks on diverse foundation LLMs.
Based on the findings obtained from the empirical study, 
% We propose an adaptive planning framework named AdaCoder,
% It initializes a non-planning code generation first. The LLM-based generation is followed by lightweight script-based testing and rule-based debugging, avoiding the complex collaboration and high computational cost of LLM-based agents. If the debugged code still fails to pass the given test cases, an LLM-based planning agent generates a planning prompt that adapts to error feedback and triggers a regeneration. This iteration takes advantage of the complementary strengths between planning- and non-planning-based generation.
we propose AdaCoder, a novel adaptive planning, multi-agent
framework for function-level code generation. AdaCoder has two phases. Phase-1 is an initial code generation step without planning, which uses an LLM-based coding agent and a script-based testing agent to unleash LLM's native power, identify cases beyond LLM's power, and determine the errors hindering execution. Phase-2 adds a rule-based debugging agent and an LLM-based planning agent for iterative code generation with planning. %The debugging agent adaptively fixes superficial errors using a rule-based method, while the planning agent guides the coding agent to address in-depth errors by adaptively generating a step-by-step plan explaining past failures and correct solutions based on error feedback.
% involves initial code generation without planning, while Phase-2 employs iterative code generation with planning. This approach leverages lightweight script-based testing and rule-based debugging adapted to three types of common errors, avoiding the complex collaboration and high computational cost of LLM-based agents. If the debugged code still fails to pass the given test cases, an LLM-based planning agent generates a planning prompt that adapts to error feedback and triggers a regeneration for another iteration, taking advantage of the complementary strengths between planning- and non-planning-based generation.
Our evaluation shows that AdaCoder achieves higher generalizability on diverse LLMs. Compared to the best baseline MapCoder, AdaCoder is on average 27.69\% higher in Pass@1, 16 times faster in inference, and 12 times lower in token consumption.

% \fi
\IEEEpeerreviewmaketitle

\section{Introduction} \label{intro}

Code generation refers to the automatic translation of natural language descriptions of software development tasks into code snippets written in a programming language %with correct functionality 
\cite{poesia2022synchromesh}. Intelligent programming assistants such as GitHub Copilot \cite{chen2021evaluatinglargelanguagemodels} can help developers reduce their programming efforts in writing repetitive code functions and improve their development productivity \cite{xu2020incorporating, guo2020graphcodebert}. Essentially, these assistants are powered by Large Language Models (LLMs), such as Codex \cite{chen2021evaluatinglargelanguagemodels} and DeepSeek-Coder \cite{guo2024deepseek}. %These foundation LLMs were pre-trained on extensive codebases and fine-tuned specifically for code generation tasks. 

To evaluate the performance of LLMs, researchers have developed various benchmarks, which can be categorized into function-level (e.g., HumanEval \cite{chen2021evaluatinglargelanguagemodels}), class-level (e.g., ClassEval \cite{du2023classevalmanuallycraftedbenchmarkevaluating}) and repository-level (e.g., RepoCoder \cite{zhang2023repocoder}). 
Function-level benchmarks assess the functional correctness of the generated code, while class- and repository-level benchmarks evaluate whether a generated code matches the context requirement of a given class or repository. This paper focuses on the function level, which satisfy developers' fine-grained programming requirements \cite{chen2021evaluatinglargelanguagemodels, austin2021programsynthesislargelanguage}, supports many intelligent programming assistants \cite{Wong_2023, Corso_2024}, and serves as a research foundation for class- and repository-level generation \cite{zhang2023repocoder}.

Recently, researchers presented the multi-agent framework for the generation of function-level code \cite{huang2024agentcodermultiagentbasedcodegeneration, islam2024mapcodermultiagentcodegeneration, wang2024intervenorpromptingcodingability, dong2024self, hong2023metagptmetaprogrammingmultiagent}. The multi-agent framework first leverages an LLM (e.g., GPT-4) as an independent agent serving for a programming task, such as task planning, code generation, test case generation, or bug repairing \cite{dong2024self}. The framework then determines the workflow for collaboration of several included agents \cite{islam2024mapcodermultiagentcodegeneration}. 
For example, Self-Collaboration proposed by Dong et al. \cite{dong2024self} uses GPT-4 \cite{achiam2023gpt} and GPT-3.5 \cite{brown2020language} as the foundation LLM, enhancing code generation capabilities at the function level by forming a team of agents for collaborative work. MapCoder presented by Islam et al. \cite{islam2024mapcodermultiagentcodegeneration} consists of four LLM-based agents that simulate the entire cycle of human developers writing code, greatly enhancing the performance of code generation. 

However, while these frameworks \cite{huang2024agentcodermultiagentbasedcodegeneration, islam2024mapcodermultiagentcodegeneration, wang2024intervenorpromptingcodingability, dong2024self, hong2023metagptmetaprogrammingmultiagent} perform well on ChatGPT series models, %the lack of disclosed implementation details or performance results 
it is unclear how they can be generalized to other foundation LLMs with varying parameter sizes, architectures, and performance levels. % makes their generalizability unclear.
To address this, we conduct an empirical study to investigate the generalizability of existing multi-agent frameworks. Specifically, we evaluate four state-of-the-art multi-agent frameworks (AgentCoder \cite{huang2024agentcodermultiagentbasedcodegeneration}, MapCoder \cite{islam2024mapcodermultiagentcodegeneration}, INTERVENOR \cite{wang2024intervenorpromptingcodingability}, and Self-Collaboration \cite{dong2024self}) using six diverse open-source LLMs (CodeLlama-Python 7B/13B/34B \cite{rozière2024codellamaopenfoundation} and DeepSeek-Coder 1.3B/6.7B/33B \cite{guo2024deepseek}) on the widely used HumanEval benchmark \cite{chen2021evaluatinglargelanguagemodels}.
% In this work, we conduct an empirical study on the generalizability of state-of-the-art, multi-agent code generation frameworks, including AgentCoder \cite{huang2024agentcodermultiagentbasedcodegeneration}, MapCoder \cite{islam2024mapcodermultiagentcodegeneration}, INTERVENOR \cite{wang2024intervenorpromptingcodingability}, and Self-Collaboration \cite{dong2024self}. We tested their generalizaiblity with six open-source LLMs (CodeLlama-Python 7B/13B/34B \cite{rozière2024codellamaopenfoundation} and DeepSeek-Coder 1.3B/6.7B/33B \cite{guo2024deepseek}) on the widely used HumanEval benchmark \cite{chen2021evaluatinglargelanguagemodels}. 
Compared to the ChatGPT series models with more than 175B parameters \cite{floridi2020gpt} and Pass@1 results ranging from 60.3\% to 90.2\% \cite{gpt4result}, these LLMs span relatively smaller parameter scales (1.3B to 34B), have different architectures, and exhibit a wide range of lower performance levels (Pass@1 from 32.69\% to 64.63\%). The empirical study showed that:
\begin{itemize}
    \item[$\bullet$] MapCoder achieves the best results because of its LLM-based task planning, but at an extremely high inference cost; non-planning generation is highly complementary to planning.
    \item[$\bullet$] INTERVENOR’s collaboration among LLM-based agents is relatively simple, providing consistent performance improvements for base models, but still falls short of MapCoder.
    \item[$\bullet$] AgentCoder and Self-Collaboration show poor generalizability with degraded performance. The iterative workflow of all frameworks is ineffective.\vspace{3pt}
\end{itemize}

% AgentCoder, INTERVENOR, and Self-Collaboration perform poorly and may even degrade performance. Only MapCoder demonstrates substantial enhancement for open-source LLMs, but with high token consumption and long inference time. The effectiveness attributes to its planning mechanism, i.e., Multi-Plan Coding, which guide LLMs in generations with various plans. An in-depth analysis indicates that combining planning and non-planning mechanisms properly shows potential for better performance and less computational resources. Besides, the other frameworks failed due to the ineffective iterations of multi-agent collaborations.

Based on our empirical findings, we designed an adaptive planning framework for multi-agent code generation called \textbf{AdaCoder}. 
The framework has two phases. Phase-1 is an initial code generation without planning, using an LLM-based agent \textit{Programming Assistant} for coding and a script-based agent \textit{Code Evaluator} for testing. This phase unleashes the LLM's native power and identify cases beyond LLM's power or errors hindering execution. To reduce inference cost and ensure accurate information transfer, the \textit{Code Evaluator} assesses code correctness using the sample test cases given in the task description, instead of using LLM-based test case generation as AgentCoder does. Phase-2 adds a rule-based debugging agent \textit{Debug Specialist} and an LLM-based planning agent \textit{Prompt Engineer} for iterative code generation with planning. The \textit{Debug Specialist} adaptively fixes superficial errors using a rule-based method, derived from our prior work \cite{wen2024fixingcodegenerationerrors}, based on the error feedback from the \textit{Code Evaluator}. This replaces costly LLM-based bug localization \cite{qin2024agentfl} and debugging \cite{lee2024unified}. The \textit{Prompt Engineer} guides the Programming Assistant to address in-depth errors by adaptively generating a step-by-step plan explaining past failures and correct solutions based on error feedback.
% We propose AdaCoder, an adaptive planning framework with two phases. Phase-1 is an initial code generation without planning, using a coding agent and a testing agent to unleash LLM's native power and identify cases beyond LLM's power or hindering execution. Phase-2 adds a debugging agent and a planning agent for iterative code generation with planning to address superficial and in-depth errors. The debugging agent fixes superficial errors using a rule-based method, while the planning agent guides the coding agent to address in-depth errors by generating a step-by-step plan explaining past failures and correct solutions.
% Generally, AdaCoder consists of four agents: Programming Assistant, an LLM-based agent that generates code for a given task prompt; Debug Specialist, a rule-based method based on previous work \cite{wen2024fixingcodegenerationerrors} for fixing compilation and runtime errors in generated code; Code Evaluator, a script capable of interacting with the local development environment and collect feedback from Programming Assistant and Debug Specialist; Prompt Engineer, an LLM-driven agent that generates a new generation plan if the generation error cannot be fixed to pass the Code Evaluator's testing, and guide the Programming Assistant for code generation from scratch. The initial non-planning generation plus rule-based debugging provides a solution for simple programming tasks. The simplified iteration fully makes use of the planning mechanism when the non-planning fails, without incorporating noisy information and high computation cost.

To evaluate the effectiveness and generalizability of AdaCoder, we applied it to six different LLMs used in our empirical study and four ChatGPT series LLMs (i.e., GPT-3.5-turbo, GPT-4, GPT-4-turbo, and GPT-4o). We tested them on HumanEval %and another widely used function-level 
and MBPP \cite{austin2021programsynthesislargelanguage} code generation benchmarks.
%MBPP \cite{austin2021programsynthesislargelanguage}. 
The experimental results show that AdaCoder can effectively improve the performance of all LLMs with high generalizability, and can significantly outperform the best baseline MapCoder by 27.69\% on average in terms of Pass@1. Moreover, AdaCoder achieves 16X faster inference speed and 12X less token consumption than MapCoder.

In summary, the major contributions of this paper are as follows:

\begin{itemize}
    % \item[$\bullet$] Conducting an empirical study on representative multi-agent frameworks on two families of open-source LLMs.
    
    \item[$\bullet$] Evaluating the generalizability of four state-of-the-art multi-agent frameworks across six different LLMs, covering a wide spectrum of parameter scales and performance.\vspace{3pt}
    
    % \item[$\bullet$] , such as noise and the planning mechanism, that affect the performance of multi-agent frameworks.
    
    \item[$\bullet$] Identifying the influential factors on the design of multi-agent frameworks and proposing an adaptive planning framework AdaCoder \cite{replication2024package}.\vspace{3pt}
    
    \item[$\bullet$] Demonstrating AdaCoder's generalizability across LLMs with varying parameter sizes, architectures, and performance levels, while showcasing its high effectiveness and low computational cost over baselines.
\end{itemize}

\section{Related Work} \label{related}
% This section describes the existing code generation studies in subsection \ref{code_generation}. It then provides the relevant methods for enhancing code generation capabilities in subsection \ref{enhancing_code_generation}, leading to the goal of this study.

% \subsection{LLMs for Function-Level Code Generation} \label{code_generation}
% \vspace{5pt}\noindent\textbf{LLMs for Function-Level Code Generation.}
\subsection{LLMs for Function-Level Code Generation}
Code generation \cite{chen2021evaluatinglargelanguagemodels, austin2021programsynthesislargelanguage, zheng2024codegeexpretrainedmodelcode} is an automated process for producing program code, aiming to reduce manual coding efforts and improve software development efficiency. Code generation methods cover many techniques \cite{puschel2005spiral, yin2017syntactic, syriani2018systematic, svyatkovskiy2020intellicode, rozière2024codellamaopenfoundation, danilchenko2012automated}, ranging from template-based approaches \cite{danilchenko2012automated,bajwa2006rule,van2018automated} to deep learning models capable of comprehending complex requirements \cite{le2022coderl,sutskever2014sequence,Roziere2020TransCoder}. Recently, LLMs have brought new breakthroughs in code generation tasks. %, with GPT \cite{radford2018improving} and BERT \cite{devlin2018bert} as typical representatives. 
%LLMs such as CodeLlama \cite{rozière2024codellamaopenfoundation} and DeepSeek-Coder \cite{guo2024deepseek} are artificial intelligence systems based on deep learning, trained on massive text data, containing billions to trillions of parameters and possessing powerful language understanding and generation capabilities. 
Compared to traditional models, LLMs can better understand the programming requirements expressed by developers and generate more satisfactory code \cite{liu2023improving}. Currently, various LLMs have been developed for code generation tasks \cite{li2022competition, fried2023incodergenerativemodelcode, chen2021evaluatinglargelanguagemodels, austin2021programsynthesislargelanguage, nijkamp2023codegenopenlargelanguage, allal2023santacoderdontreachstars, rozière2024codellamaopenfoundation, gunasekar2023textbooksneed, li2023textbooksneediiphi15, achiam2023gpt}. 
\subsection{Prompt Engineering for LLM-based Code Generation}
Although there are numerous LLMs with robust code generation capabilities such as CodeLlama \cite{rozière2024codellamaopenfoundation} and DeepSeek-Coder \cite{guo2024deepseek} currently exist \cite{zheng2023survey}, there remains a significant potential for further performance enhancement \cite{achiam2023gpt}
% . Such improvements can be achieved not only through fundamental architectural refinements of the LLMs themselves but also through some innovative method, 
including Prompt Engineering and Iterative Refinement. Prompt engineering is an advanced technique that optimizes LLM's output through meticulously designed input prompts. Liu et al. \cite{liu2023improving} discovered that LLMs' performance is highly sensitive to prompts, particularly to programming requirements expressed in natural language. % as model input. 
Wei et al. \cite{wei2022chain} introduced the innovative Chain-of-Thought prompting technique, which substantially improved the performance of LLM in reasoning tasks and code generation. Yao et al. \cite{yao2023treethoughtsdeliberateproblem} proposed the Tree-of-Thought (ToT) method, which enables LLMs to engage in deliberate decision making by considering multiple reasoning paths and self-evaluating choices to determine the next course of action, thus significantly enhancing LLMs' capability to solve complex tasks. 
% Zhang et al. \cite{zhang2024cumulativereasoninglargelanguage} introduced Cumulative Reasoning (CR), a novel approach that utilizes language models cumulatively and iteratively, mirroring human thought processes for problem-solving. CR decomposes tasks into smaller, manageable components and leverages previous propositions for effective composition, significantly enhancing problem-solving capabilities. 

% \vspace{5pt}\noindent\textbf{Iteration Refinement for LLMs.}
\subsection{Iterative Refinement for LLM-based Code Generation}
Iterative refinement is an advanced method that progressively enhances the quality of the code through multiple modifications. This approach ingeniously mimics the process of human programmers writing and debugging code. Olausson et al. \cite{olausson2023self} conducted an in-depth study on the effectiveness of Self-Repair in code generation tasks. Their experimental results indicate that Self-Repair can, to a certain extent, improve the quality of code generated by LLMs, albeit at the cost of increased GPU resources and time investment. Chen et al. \cite{chen2023teachinglargelanguagemodels} proposed the innovative Self-Debug method, which teaches LLMs how to debug their generated code through examples, achieving state-of-the-art performance across multiple code generation benchmarks. Zhang et al. \cite{zhang-etal-2023-self} introduced the Self-Edit technique, which cleverly utilizes the execution results of LLM-generated code to enhance its performance in code generation tasks. This method achieved significant performance improvements ranging from 31\% to 89\% on nine code-generating LLMs and three code-generation benchmarks. 

% \subsection{LLM-Based Multi-Agent Framework}
% \vspace{5pt}\noindent\textbf{LLM-Based Multi-Agent Framework.}
\subsection{LLM-Based Multi-Agent Framework}
Multi-agent frameworks have emerged as an innovative approach in recent months, ingeniously simulating the process of multiple experts collaborating to solve problems. Huang et al. \cite{huang2024agentcodermultiagentbasedcodegeneration} proposed the AgentCoder framework, which comprises three LLM-based agents and utilizes test cases generated by LLM agents to improve reasoning performance. Islam et al. \cite{islam2024mapcodermultiagentcodegeneration} designed the MapCoder framework, 
which consists of four LLM agents and a planning mechanism. This mechanism first instructs the LLM to generate five distinct problems, then creates a step-by-step plan for each problem, and finally generates corresponding code based on these diverse plans. This mechanism is referred to as ``multi-plan coding" in this paper. %later discussions and 
Multi-plan coding achieves state-of-the-art results across multiple code generation benchmarks.
% consisting of four LLM agents that simulate the complete cycle of human developers writing code, achieving state-of-the-art results across multiple code generation benchmarks. 
Wang et al. \cite{wang2024intervenorpromptingcodingability} introduced INTERVENOR (INTERactiVE chaiN Of Repair), an innovative method that simulates the human process of writing and debugging code, incorporating two LLM agents and significantly improving LLM performance in code generation and translation tasks. Dong et al. \cite{dong2024self} proposed a three-LLM agent collaboration framework called Self-Collaboration, where each LLM assumes a different role, forming a virtual team for collaborative work and %ultimately 
significantly improving code generation performance. %pass@1. 
However, these studies only evaluated their effectiveness using %current state-of-the-art 
the ChatGPT series as foundation models.%, overlooking the vast community of open-source LLMs. Therefore, this research will focus on enhancing code generation capabilities for both close-source LLMs and numerous open-source LLMs. 

% Moreover, researchers almost exclusively use the GPT series models as the foundation models for multi-agent frameworks, while largely overlooking the vast community of open-source models. This is precisely the focus of this study.

\section{Empirical Study} \label{exp}
We perform an empirical study to evaluate the generalizability of state-of-the-art multi-agent frameworks across diverse LLMs with varying parameter sizes, architectures, and performance levels. We also analyze the influential factors on the performance, thereby providing insights for designing a multi-agent framework with higher generalizability, faster inference time, and less token consumption.

\subsection{Research Questions}

This empirical study aims to investigate the following two research questions (RQs).

% \vspace{0.2cm}
% \begin{mdframed}[nobreak=true]
% \textbf{RQ1: How is the generalizability of the state-of-the-art multi-agent frameworks on open-source LLMs?\hy{maybe no box for RQ questions, can only give boxes to answers}}
% \end{mdframed}
% \vspace{0.2cm}

\textbf{RQ1: How is the generalizability of the state-of-the-art multi-agent frameworks on diverse LLMs?} Existing multi-agent frameworks \cite{huang2024agentcodermultiagentbasedcodegeneration, islam2024mapcodermultiagentcodegeneration, wang2024intervenorpromptingcodingability, dong2024self, hong2023metagptmetaprogrammingmultiagent} only demonstrated effectiveness when applied to %high generation performance 
ChatGPT series LLMs, while their effectiveness on other LLMs with varying parameter sizes, architectures, and performance levels remains unexplored. %If proven ineffective, their generalizability would be significantly compromised.

% However, they predominantly rely on closed-source ChatGPT series as their foundational models, making the effectiveness of multi-agent frameworks with open-source LLMs uninvestigated. \zyh{If they prove ineffective, their generalizability will be called into question and their effectiveness is highly depended on the capability of the foundation LLMs.} 
% Compared to closed-source models like ChatGPT, open-source models generally have smaller parameter sizes, lower computational costs \cite{brown2020language}, and constitute the majority of LLMs, making them more representative.
% Compared to closed-source LLMs like ChatGPT, open-source LLMs generally have lower development and usage costs \cite{gpt4result}. Moreover, the code and training details of open-source LLMs are fully accessible \cite{guo2024deepseek}, allowing users to verify the fairness, security, and reliability of the model, or customize the architecture, training data, and parameters based on specific needs \cite{DeepSeekGGUF}, making them more representative.

% Therefore, we propose the following research question (RQ).

% \vspace{0.2cm}
% \begin{mdframed}[nobreak=true]
% \textbf{RQ2: What factors influence the effectiveness of multi-agent frameworks?}
% \end{mdframed}
% \vspace{0.2cm}

\textbf{RQ2: What factors influence the effectiveness of multi-agent frameworks?} By combining the evaluation results of these multi-agent frameworks using LLMs with varying parameter sizes, architectures, and performance levels, we aim to identify the factors that affect the performance of these frameworks. This can shed light into the development of new, effective, and efficient multi-agent frameworks.

\subsection{Dataset and Evaluation Measure}

For our empirical study, we used the HumanEval dataset \cite{chen2021evaluatinglargelanguagemodels}, a benchmark developed by OpenAI and widely used to evaluate LLMs \cite{rozière2024codellamaopenfoundation,guo2024deepseek}. This dataset comprises 164 Python programming challenges, each accompanied by an average of 7.7 test cases to verify functional correctness. 
% It has been widely adopted by prominent LLMs such as CodeLlama  and DeepSeek-Coder \cite{} to assess code generation capabilities. 
HumanEval consists of five key components: task\_id, a unique identifier for each challenge; prompt, a textual description of the generation requirements; canonical\_solution, the standard solution for the task; test, a function with multiple test cases to assess the generated code's compliance with requirements; entry\_point, the name of the main function to be generated.

% \begin{itemize}
%     \item[$\bullet$] \textbf{task\_id}: A unique identifier for each challenge.
%     \item[$\bullet$] \textbf{prompt}: A textual description of the generation requirements.
%     \item[$\bullet$] \textbf{canonical\_solution}: The standard solution for the task.
%     \item[$\bullet$] \textbf{test}: A function with multiple test cases to assess the generated code's compliance with requirements.
%     \item[$\bullet$] \textbf{entry\_point}: The name of the main function to be generated.
% \end{itemize}

% \begin{figure}
%     \centering
%     \includegraphics[width=0.8\linewidth]{image/humaneval_sample.pdf}
%     \caption{A schematic diagram of the first sample in HumanEval.}
%     \label{humaneval_sample}
% \end{figure}

On this dataset, an LLM is required to produce functionally correct code that passes all test cases based on the given prompt. To evaluate the performance of code generation, Chen et al. \cite{chen2021evaluatinglargelanguagemodels} introduced the metric $Pass@k$, which represents the percentage of tasks successfully solved by an LLM. A task is considered solved if any of the top-$k$ generated code samples pass all test cases. To address the high variance associated with Pass@k, Chen et al. \cite{chen2021evaluatinglargelanguagemodels} presented an unbiased version. Our study takes Chen et al.'s version as the evaluation measure.

%specifically focuses on $Pass@1$, as it aligns with the developers' expectations for LLMs to generate correct code on the first attempt.

% : $Pass@k=\mathbb{E}\left[1-\binom{n-c}{k}/\binom{n}{k}\right]$, where $\mathbb{E}$ denotes the average performance across all tasks; $n$ is the total number of tasks; $c$ is the number of correctly solved tasks; $\binom{n}{k}$ represents the number of combinations to choose $k$ tasks out of $n$; $\binom{n-c}{k}$ indicates the number of combinations to select $k$ tasks from those the model failed to solve. 

The inference cost is measured by the token consumption and inference time. %Both reflect the computational and resource efficiency of the framework. 
Token consumption serves as an indicator of the LLM's input-output complexity, as it determines the amount of data processed during each interaction with the LLM. Lower token consumption not only reduces the cost of API calls %(for closed-source models) 
but also decreases memory and processing overhead for open-source models. Inference time measures the framework's speed and responsiveness, which directly affects the quality of user experience.

% \begin{itemize}
%     \item[$\bullet$] $\mathbb{E}$ denotes the average performance across all tasks
%     \item[$\bullet$] $n$ is the total number of tasks
%     \item[$\bullet$] $c$ is the number of correctly solved tasks
%     \item[$\bullet$] $\binom{n}{k}$ represents the number of combinations for choosing $k$ tasks out of $n$
%     \item[$\bullet$] $\binom{n-c}{k}$ indicates the number of combinations for selecting $k$ tasks from those the model failed to solve
% \end{itemize}

\subsection{Baselines (LLM-Based Multi-Agent Frameworks)} \label{multi_agent_framework}

We leveraged four state-of-the-art multi-agent frameworks %to evaluate their performance 
as our baselines. %Generally, 
AgentCoder \cite{huang2024agentcodermultiagentbasedcodegeneration} and MapCoder \cite{islam2024mapcodermultiagentcodegeneration} are two top-performing multi-agent frameworks \cite{HumanEvalBenchmark}, while INTERVENOR \cite{wang2024intervenorpromptingcodingability} and Self-Collaboration \cite{dong2024self} represent the latest peer-reviewed state-of-the-art. All these multi-agent frameworks are reproduced by using the replication packages provided by their original studies under default settings. Specifically, 
\textit{1) AgentCoder} \cite{huang2024agentcodermultiagentbasedcodegeneration} employs GPT-4 and GPT-3.5 as the foundation LLMs, featuring three agents: Programmer for generating or repairing code, Test Designer for creating test cases, and Test Executor for evaluating code. If tests fail, error feedback is sent to the Programmer for regeneration. \textit{2) MapCoder} \cite{islam2024mapcodermultiagentcodegeneration}, utilizing GPT-4 as its foundation LLM, consists of four agents: Retrieval Agent for generating $t$ similar questions based on the original problem description, Planning Agent for creating a plan for each question and assigning confidence scores, Coding Agent for converting the highest-confidence plan into code, and Debugging Agent for debugging the code up to $k$ attempts. If debugging fails after $k$ attempts, the process is reverted to the Planning Agent to select the next highest-confidence plan. With $t$ plans in total, the entire workflow iterates up to $t$ times, resulting in a complexity of $O(kt)$. \textit{3) INTERVENOR} \cite{wang2024intervenorpromptingcodingability}, based on GPT-3.5, uses two agents: Code Learner for generating and fixing code and Code Teacher for providing repair feedback. \textit{4) Self-Collaboration} \cite{dong2024self}, also using GPT-3.5, features Analyst for decomposing tasks and creating plans, Coder for implementing solutions, and Tester for evaluating code and providing feedback. In all frameworks, iterative cycles refine code until success or maximum attempts are reached.

\subsection{Foundation LLMs} \label{exp_llm}

To comprehensively evaluate the performance of multi-agent frameworks, we have carefully selected six diverse open-source LLMs %from two series 
with varying parameter scales and performance characteristics % to serve 
as foundation LLMs. These models are as follows. 

% \begin{itemize}
    % \item[$\bullet$] 
\vspace{5pt}\noindent\textbf{CodeLlama} \cite{roziere2023code} was introduced by Meta AI.
% in August 2023, 
% This is an open-source family of code-oriented LLMs comprising three distinct model styles, reflecting the code generation capabilities of early open-source large models \cite{zheng2023survey}. In particular, 
The CodeLlama-Python series is specifically optimized for Python code generation, offering four parameter scales: 7B, 13B, and 34B. We reproduced their Pass@1 performance on the HumanEval dataset,
% , setting the temperature to 0.2 and the maximum generation length to 512 tokens, 
achieving 32.69\%, 36.65\%, and 43.72\%, respectively. Considering hardware resource constraints, we excluded the 70B version for our experiments.

    % \item[$\bullet$] 
\vspace{5pt}\noindent\textbf{DeepSeek-Coder} \cite{guo2024deepseek} is proposed by DeepSeek.
% in February 2024, 
This open-source code LLM family includes two series: base and instruct, each with three parameter specifications: 1.3B, 6.7B, and 33B. We chose the instruct series for our experiments because of its superior performance. For simplicity, we will refer to the instruct version as DeepSeek-Coder hereafter. 
% These models reflect the code generation capabilities of recent open-source code LLMs \cite{ren2024reflectioncoderlearningreflectionsequence}.
We reproduced their Pass@1 performance on the HumanEval dataset, 
% setting the temperature to 0.2 and the maximum generation length to 512 tokens, 
yielding results of 49.39\%, 58.54\%, and 64.63\% for the 1.3B, 6.7B, and 33B versions, respectively. 
% \end{itemize}

\vspace{5pt}
By selecting these six LLMs from two series, our research encompasses a wide range of parameter scales from 1.3B to 34B, including small (1.3B), medium (6.7B-13B) and large (33B-34B) models. The performance levels of these models span from 32.69\% to 64.63\%, providing us with a wide  spectrum performance. Furthermore, the CodeLlama and DeepSeek-Coder series exhibit significant differences in architecture and training methodologies, where CodeLlama is pre-trained based on the Llama2 architecture, while DeepSeek-Coder is trained from scratch using a high-quality, project-level code corpus. The diverse foundation LLMs allow for a comprehensive analysis of multi-agent framework generalizability across varying architectures, complexities, and capabilities.
% , making this comprehensive study yield reliable conclusions, reveal framework applicability and limitations, and inform new framework designs. 
All open-source LLMs are tested using HuggingFace parameters on a server with four NVIDIA RTX3090 GPUs under default settings.

% This diversity in parameter scale, performance, and architecture enables us to conduct an in-depth analysis of the adaptability of multi-agent frameworks across models with varying architectures, computational complexities, and capability levels. Such a comprehensive analysis allows us to draw more reliable conclusions, understand the framework's applicability and potential limitations, and provide broader insights for designing new multi-agent frameworks. All open-source LLMs are reproduced by using the LLM parameters stored in the HuggingFace platform and tested on a server with four NVIDIA RTX3090 GPUs in default settings. 

\subsection{Generalizability of Multi-Agent Frameworks on Open-Source LLMs (RQ1)} \label{rq1}
\subsubsection{Pass@1 Analysis of Code Generation} \label{accuracy_performance} 

\begin{table*}
    \centering
    \scriptsize
    \caption{Pass@1 performance of multi-agent frameworks combined with the six selected LLMs on the HumanEval benchmark. ``Direct" refers to instructing LLMs to generate code without the use of any multi-agent framework, relying solely on the inherent capabilities of the LLM. In subsequent tables, ``Direct" consistently carries this meaning.}\label{tab_overview_a}
    \setlength{\tabcolsep}{11.5pt}{
    \begin{tabular}{lccccc}
        \toprule
        \textbf{LLMs} & \textbf{Direct} & \textbf{AgentCoder} & \textbf{MapCoder} & \textbf{INTERVENOR} & \textbf{Self-Collaboration} \\
        \midrule
        CodeLlama-Python-7B & 32.69 & 31.71 (\textcolor{red}{$\downarrow$03.00\%}) & 52.44 (\textcolor{cgreen}{$\uparrow$60.42\%}) & 40.85 (\textcolor{cgreen}{$\uparrow$24.96\%}) & 31.71 (\textcolor{red}{$\downarrow$03.00\%}) \\
        CodeLlama-Python-13B & 36.65 & 35.98 (\textcolor{red}{$\downarrow$01.83\%}) & 59.15 (\textcolor{cgreen}{$\uparrow$61.39\%}) & 48.17 (\textcolor{cgreen}{$\uparrow$31.43\%}) & 41.46 (\textcolor{cgreen}{$\uparrow$13.12\%}) \\
        CodeLlama-Python-34B & 43.72 & 48.17 (\textcolor{cgreen}{$\uparrow$10.18\%}) & 73.78 (\textcolor{cgreen}{$\uparrow$68.76\%}) & 56.10 (\textcolor{cgreen}{$\uparrow$28.32\%}) & 44.51 (\textcolor{cgreen}{$\uparrow$01.81\%}) \\
        DeepSeek-Coder-1.3B & 49.39 & 34.15 (\textcolor{red}{$\downarrow$30.86\%}) & 60.37 (\textcolor{cgreen}{$\uparrow$22.23\%}) & 53.05 (\textcolor{cgreen}{$\uparrow$07.41\%}) & 42.07 (\textcolor{red}{$\downarrow$14.82\%}) \\
        DeepSeek-Coder-6.7B & 58.54 & 60.37 (\textcolor{cgreen}{$\uparrow$03.13\%}) & 67.68 (\textcolor{cgreen}{$\uparrow$15.61\%}) & 67.68 (\textcolor{cgreen}{$\uparrow$15.61\%}) & 40.85 (\textcolor{red}{$\downarrow$30.22\%}) \\
        DeepSeek-Coder-33B & 64.63 & 46.95 (\textcolor{red}{$\downarrow$27.36\%}) & 67.68 (\textcolor{cgreen}{$\uparrow$04.72\%}) & 67.07 (\textcolor{cgreen}{$\uparrow$03.78\%}) & 45.12 (\textcolor{red}{$\downarrow$30.19\%}) \\
        \midrule
        $\Delta$ Average & - & \textcolor{red}{$\downarrow$08.29\%} & \textcolor{cgreen}{$\uparrow$38.86\%} & \textcolor{cgreen}{$\uparrow$18.59\%} & \textcolor{red}{$\downarrow$10.55\%} \\
        \bottomrule
    \end{tabular}}
\end{table*}

\begin{table*}
    \centering
    \scriptsize
    \caption{Average token consumption for inference by multi-agent frameworks with the six selected LLMs on the HumanEval benchmark, "$\times$" represents multiple of increase.}\label{tab_overview_b}
    \setlength{\tabcolsep}{10pt}{
    \begin{tabular}{lccccc}
        \toprule
        \textbf{LLMs} & \textbf{Direct} & \textbf{AgentCoder} & \textbf{MapCoder} & \textbf{INTERVENOR} & \textbf{Self-Collaboration} \\
        \midrule
        CodeLlama-Python-7B & 00.91K & 13.15K (\textcolor{cgreen}{$\uparrow$13.47$\times$}) & 24.72K (\textcolor{cgreen}{$\uparrow$26.20$\times$}) & 16.59K (\textcolor{cgreen}{$\uparrow$17.25$\times$}) & 27.94K (\textcolor{cgreen}{$\uparrow$29.73$\times$}) \\
        CodeLlama-Python-13B & 00.92K & 12.72K (\textcolor{cgreen}{$\uparrow$12.81$\times$}) & 21.79K (\textcolor{cgreen}{$\uparrow$22.66$\times$}) & 14.33K (\textcolor{cgreen}{$\uparrow$14.56$\times$}) & 21.19K (\textcolor{cgreen}{$\uparrow$22.01$\times$}) \\
        CodeLlama-Python-34B & 00.95K & 10.91K (\textcolor{cgreen}{$\uparrow$10.45$\times$}) & 19.72K (\textcolor{cgreen}{$\uparrow$19.69$\times$}) & 11.20K (\textcolor{cgreen}{$\uparrow$10.75$\times$}) & 19.87K (\textcolor{cgreen}{$\uparrow$19.85$\times$}) \\
        DeepSeek-Coder-1.3B & 01.13K & 12.82K (\textcolor{cgreen}{$\uparrow$10.36$\times$}) & 27.30K (\textcolor{cgreen}{$\uparrow$23.20$\times$}) & 08.26K (\textcolor{cgreen}{$\uparrow$06.33$\times$}) & 18.32K (\textcolor{cgreen}{$\uparrow$15.24$\times$}) \\
        DeepSeek-Coder-6.7B & 01.16K & 09.62K (\textcolor{cgreen}{$\uparrow$07.29$\times$}) & 29.92K (\textcolor{cgreen}{$\uparrow$24.77$\times$}) & 09.54K (\textcolor{cgreen}{$\uparrow$07.22$\times$}) & 22.38K (\textcolor{cgreen}{$\uparrow$18.27$\times$}) \\
        DeepSeek-Coder-33B & 01.18K & 10.88K (\textcolor{cgreen}{$\uparrow$08.23$\times$}) & 26.60K (\textcolor{cgreen}{$\uparrow$21.58$\times$}) & 10.36K (\textcolor{cgreen}{$\uparrow$07.79$\times$}) & 22.50K (\textcolor{cgreen}{$\uparrow$18.10$\times$}) \\
        \midrule
        $\Delta$ Average & - & \textcolor{cgreen}{$\uparrow$10.44$\times$} & \textcolor{cgreen}{$\uparrow$23.02$\times$} & \textcolor{cgreen}{$\uparrow$10.65$\times$} & \textcolor{cgreen}{$\uparrow$20.53$\times$} \\
        \bottomrule
    \end{tabular}}
\end{table*}

\begin{table*}
    \centering
    \scriptsize
\caption{Average inference time of multi-agent frameworks with the six selected LLMs on the HumanEval benchmark, "$\times$" represents multiple of increase.}\label{tab_overview_c}
\setlength{\tabcolsep}{10pt}{
    \begin{tabular}{lccccc}
        \toprule
        \textbf{LLMs} & \textbf{Direct} & \textbf{AgentCoder} & \textbf{MapCoder} & \textbf{INTERVENOR} & \textbf{Self-Collaboration} \\
        \midrule
        CodeLlama-Python-7B & 27.33s & 130.5s (\textcolor{cgreen}{$\uparrow$03.78$\times$}) & 851.0s (\textcolor{cgreen}{$\uparrow$30.14$\times$}) & 276.4s (\textcolor{cgreen}{$\uparrow$09.11$\times$}) & 357.8s (\textcolor{cgreen}{$\uparrow$12.09$\times$}) \\
        CodeLlama-Python-13B & 41.42s & 363.4s (\textcolor{cgreen}{$\uparrow$07.77$\times$}) & 781.6s (\textcolor{cgreen}{$\uparrow$17.87$\times$}) & 416.2s (\textcolor{cgreen}{$\uparrow$09.05$\times$}) & 387.2s (\textcolor{cgreen}{$\uparrow$08.35$\times$}) \\
        CodeLlama-Python-34B & 76.49s & 404.7s (\textcolor{cgreen}{$\uparrow$04.29$\times$}) & 865.0s (\textcolor{cgreen}{$\uparrow$10.31$\times$}) & 435.8s (\textcolor{cgreen}{$\uparrow$04.70$\times$}) & 636.5s (\textcolor{cgreen}{$\uparrow$07.32$\times$}) \\
        DeepSeek-Coder-1.3B & 22.52s & 109.3s (\textcolor{cgreen}{$\uparrow$03.85$\times$}) & 286.3s (\textcolor{cgreen}{$\uparrow$11.71$\times$}) & 135.0s (\textcolor{cgreen}{$\uparrow$05.00$\times$}) & 174.1s (\textcolor{cgreen}{$\uparrow$06.73$\times$}) \\
        DeepSeek-Coder-6.7B & 20.65s & 96.20s (\textcolor{cgreen}{$\uparrow$03.66$\times$}) & 292.7s (\textcolor{cgreen}{$\uparrow$13.18$\times$}) & 119.3s (\textcolor{cgreen}{$\uparrow$04.78$\times$}) & 223.6s (\textcolor{cgreen}{$\uparrow$09.83$\times$}) \\
        DeepSeek-Coder-33B & 88.27s & 448.9s (\textcolor{cgreen}{$\uparrow$04.08$\times$}) & 1071s (\textcolor{cgreen}{$\uparrow$11.13$\times$}) & 523.1s (\textcolor{cgreen}{$\uparrow$04.93$\times$}) & 817.6s (\textcolor{cgreen}{$\uparrow$08.26$\times$}) \\
        \midrule
        $\Delta$ Average & - & \textcolor{cgreen}{$\uparrow$04.57$\times$} & \textcolor{cgreen}{$\uparrow$15.72$\times$} & \textcolor{cgreen}{$\uparrow$06.26$\times$} & \textcolor{cgreen}{$\uparrow$08.76$\times$} \\
        \bottomrule
    \end{tabular}}
\end{table*}

Table \ref{tab_overview_a} presents the effectiveness of the four multi-agent frameworks on the HumanEval benchmark, using six selected %diverse
LLMs as foundation models.%, setting the temperature to 0.2 and the maximum generation length to 512 tokens. 

\vspace{5pt}\noindent\textbf{From the perspective of multi-agent frameworks}, both MapCoder and INTERVENOR enhance the code generation capabilities of these LLMs, but MapCoder achieves the most significant improvement, with an average increase of nearly 40\%. MapCoder's enhancement ranges from 4.72\% to 68.76\%, with DeepSeek-Coder-33B showing the smallest improvement and CodeLlama-Python-34B showing the largest, resulting in an average improvement of 38.86\%. INTERVENOR's enhancement spans from 3.78\% to 31.43\%, again with the DeepSeek-Coder-33B model showing the smallest improvement and the CodeLlama-Python-13B model showing the largest, leading to an average improvement of 18.59\%. 

In contrast, Self-Collaboration and AgentCoder tend to diminish these capabilities. Self-Collaboration's influence on selected LLM code generation performance fluctuates between -30.22\% and 13.12\%. DeepSeek-Coder-6.7B experiences the largest performance decline, while CodeLlama-Python-13B shows the most notable performance enhancement. On average, Self-Collaboration decreases the code generation pass@1 of the six LLMs 
by 10.55\%. Similarly, AgentCoder's impact on the code generation performance of the six LLMs 
ranges from -30.86\% to 10.18\%. DeepSeek-Coder-1.3B shows the largest performance decrease, while CodeLlama-Python-34B demonstrates the most significant performance improvement. On average, AgentCoder reduces the code generation capability of the six LLMs by 8.29\%. 
%Previously, we tested it on GPT-4 following Huang et al.'s \cite{huang2024agentcodermultiagentbasedcodegeneration} study and achieved a pass@1 score of 95.6\%, which is close to the original reported 96.3\%, indicating the replication correction. 
We observed that when testing AgentCoder with the six LLMs, it often generated buggy code that cannot be fixed by the subsequent agents. 
%Thus, this result implies that the six LLMs cannot understand the prompt description well due to the designed complex prompt template.

% \zyh{This is unusual because AgentCoder when based on GPT-4 for HumanEval, achieves a pass@1 score of 96.3\%, which is very high and has been reproduced by us, indicating its strong performance. However, when using AgentCoder with open-source LLMs like CodeLlama, we observed that its overly complex prompts confuse the models, leading to degraded performance.}
%\hy{according to the AgenetCoder paper, they can achieve good results with GPT-4 and GPT-3.5, and outperfom direct invocations of GPT-4. Their accuracy results on HumanEval is 96.3\%, which is very high. So are your results correct? If yes, what could be the reasons? Also, I notice that on CodeLlma34B, your result is 43.72, which is lower than the result in AgentCoder's paper (51.8). Can you reproduce AgentCoder's results with GPT-4/GPT-3.5/CodeLlma?}

% In stark contrast to the aforementioned frameworks, both MapCoder and INTERVENOR achieve improvements in code generation capabilities. 

In summary, MapCoder and INERVENOR show improvements on all selected LLMs %with generalizability
while the others do not. Also, MapCoder achieves the best performance.

% different multi-agent frameworks have significantly varying effects on enhancing the code generation capabilities of open-source LLMs, with MapCoder notably outperforming the other three multi-agent frameworks. The specific reasons underlying this phenomenon will be explored and analyzed in Section \ref{rq2}.

% \begin{figure}
%     \centering
%     \includegraphics[width=\linewidth]{image/scatter_llm.pdf}
%     \caption{A linear regression analysis using the parameter scales as the x-axis and the percentage improvement in code generation capability as the y-axis.}  
%     \label{fig_scatter_llm}
% \end{figure}

\vspace{5pt}\noindent\textbf{From the perspective of foundation LLMs}, for the CodeLlama-Python series, models with parameter scales of 7B, 13B, and 34B demonstrated average improvements in code generation capabilities of 19.85\%, 26.03\%, and 27.27\%, respectively, after applying multi-agent frameworks, showing an increasing trend with larger parameter sizes.
% We conducted a linear regression analysis using parameter count as the independent variable and percentage improvement in code generation capability as the dependent variable. The resulting linear regression equation is $y=0.2185x+20.451$, with a coefficient of determination $R^2$ of 0.6072. This indicates a 
% % relatively significant 
% positive correlation between performance improvement and parameter count for CodeLlama-Python models. 

However, we observed a %markedly 
different scenario with the DeepSeek-Coder series models. Models with parameter scales of 1.3B, 6.7B, and 33B exhibited average changes in code generation capabilities of -4.01\%, +1.03\%, and -12.27\%, respectively, after applying multi-agent frameworks. These results suggest that there is no apparent correlation between performance change and parameter scale for the DeepSeek-Coder series. Notably, the 33B model, which has the largest parameter count, showed the most substantial decrease in code generation capability after applying multi-agent frameworks, contrasting sharply with the earlier conclusion. 
% We assess the correlation between the effectiveness of frameworks and the parameter scale of LLM via linear regression. The resulting linear regression equation is $y=0.0841x+8.3185$, with a coefficient of determination $R^2$ of 0.005. This result suggests that when considering all models, there is no positive correlation between performance improvement and model parameter count. 
% , we plotted a scatter graph using the parameter scales of all six models from both series as the x-axis and the percentage improvement in code generation capability as the y-axis, followed by a linear regression analysis. As shown in Figure \ref{fig_scatter_llm}, 

In summary, the effectiveness of the multi-agent frameworks correlates to architecture of LLMs, instead of the generation capability and parameter size of LLMs.
% The above results indicate that multi-agent frameworks consistently produced positive improvements for CodeLlama-Python series, while mostly generating negative impacts on DeepSeek-Coder series, revealing a clear disparity between the two. These findings demonstrate that the effectiveness of multi-agent frameworks has no significant correlation with model parameter scale and exhibits distinct preferences for different model architectures. 

\vspace{5pt}
\subsubsection{Cost Analysis of Code Generation} \label{cost_performance} 

To analyze the resource consumption, we measured the average number of tokens consumed and inference time per sample as presented in Tables \ref{tab_overview_b}-\ref{tab_overview_c}. %The token consumption to some extent reflects GPU resource utilization, while inference time directly indicates the time cost of using a specific multi-agent framework. 

\vspace{5pt}\noindent\textbf{From the perspective of multi-agent frameworks}, AgentCoder, MapCoder, INTERVENOR, and Self-Collaboration demonstrated average increases in token consumption by factors of 10.44, 23.02, 10.65, and 20.53, respectively. Meanwhile, their average inference times increased by factors of 4.57, 15.72, 6.26, and 8.76, respectively. Among these, AgentCoder exhibited the smallest increase in both token consumption and inference time. MapCoder showed significantly higher increases in both metrics compared to the other three multi-agent frameworks. For instance, when using CodeLlama-Python-7B as the foundation model, MapCoder's inference time was 6.52 times that of AgentCoder. This phenomenon indicates that while MapCoder excels in improving code generation performance, it also incurs notably higher GPU resource and time costs. 

In summary, multi-agent frameworks generally result in a several-fold increase in inference cost, while the best state-of-the-art MapCoder exhibiting the highest increase.

% \begin{figure}
%     \centering
%     \includegraphics[width=\linewidth]{image/scatter_cost.pdf}
%     \caption{A linear regression analysis using parameter scale as the x-axis and the increase factors of token consumption and inference time as the y-axis.} % 
%     \label{fig_scatter_cost}
% \end{figure}

\vspace{5pt}\noindent\textbf{From the perspective of foundation LLMs}, after applying multi-agent frameworks, the token consumption of CodeLlama-Python 7B, 13B and 34B models increased by average factors of 21.66, 18.01, and 15.19, respectively, decreasing as parameter size grows. Similarly, their inference times increased by average factors of 13.78, 10.76, and 6.66, respectively, also showing a decreasing trend as parameter size increases.
% We conducted a linear regression analysis between token consumption and inference time
%as the y-axis, as illustrated in Figure \ref{fig_scatter_cost}. 
% The coefficient of determination $R^2$ for both fitting curves approaches 0.9, strongly indicating that for CodeLlama-Python models, the increase factors of both token consumption and inference time exhibit a negative correlation with the model's parameter scale. 
In contrast, for the DeepSeek-Coder series, after applying multi-agent frameworks, the token consumption of 1.3B, 6.7B, and 33B models increased by average factors of 13.78, 14.39, and 13.93, respectively, with a standard deviation of 0.32. Their inference times increased by average factors of 6.82, 7.86, and 7.10, respectively, with a standard deviation of 0.54. The standard deviations for both metrics are less than 10\% of their respective means, indicating low variability in the increase factors of token consumption and inference time for the DeepSeek-Coder models. This finding further corroborates that the application effects of multi-agent frameworks demonstrate distinct preferences for models with different architectures. 

In summary, the increase in inference cost is also significantly influenced by the architecture of the foundation LLMs.

\vspace{0.3cm}
\begin{mdframed}[nobreak=true]
\textbf{Answer to RQ1:} 
% MapCoder achieves the greatest improvement but at a high inference cost, followed by INTERVENOR with moderate gains, while AgentCoder and Self-Collaboration result in negative effects.
AgentCoder and Self-Collaboration yield poor generalizability on the six open-source LLMs. MapCoder achieves the best performance and good  generalizability but with high inference cost.
% generalizability
\end{mdframed}

\subsection{Factors Analysis on the Performance of Multi-Agent Frameworks (RQ2)} \label{rq2}

\vspace{5pt}
\subsubsection{Effectiveness Analysis of Iterative Refinement} \label{iteration_refinement}

\begin{figure*}
    \centering
    \includegraphics[width=1\linewidth]{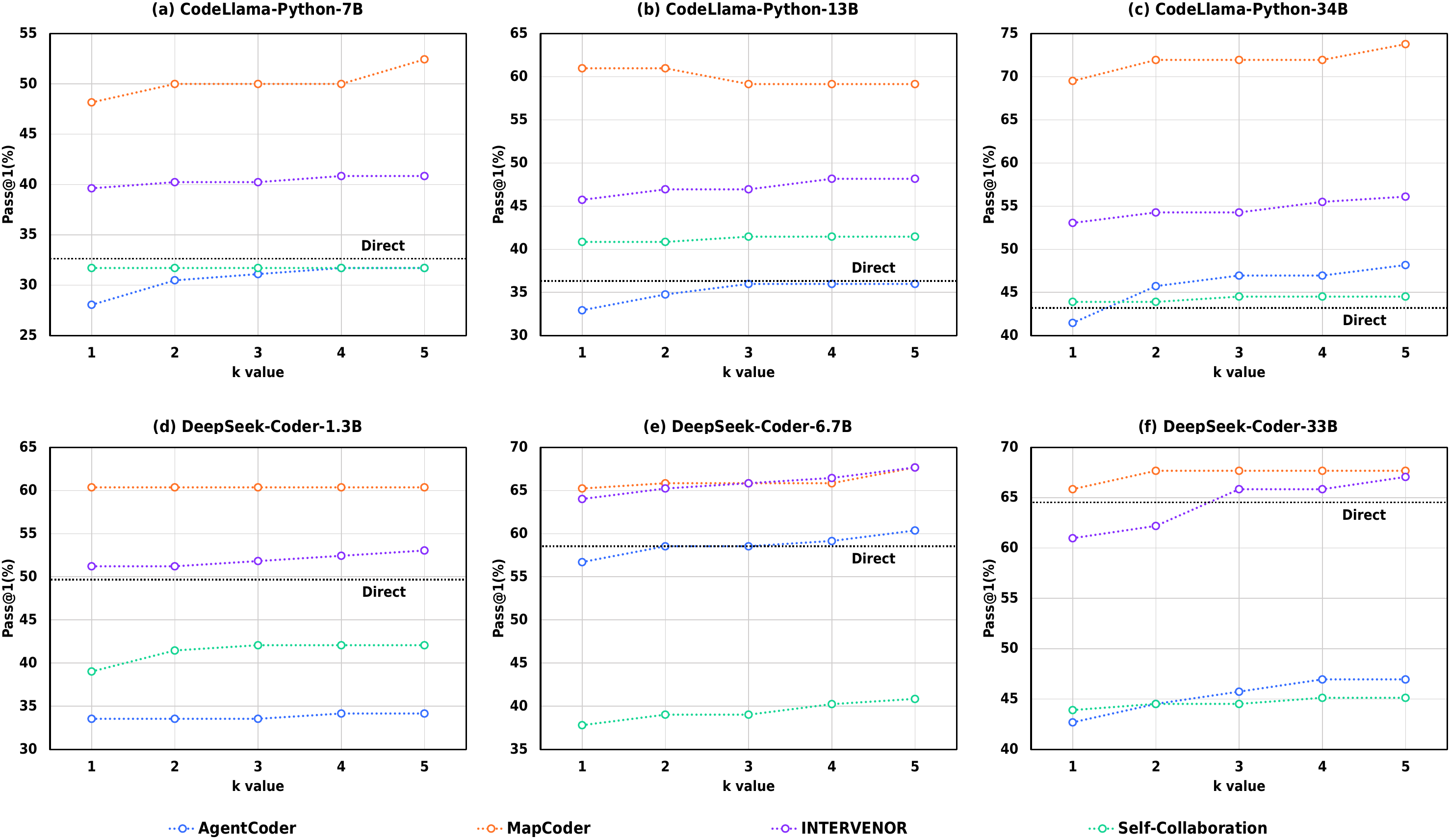}
    \caption{Line graphs of pass@1 for six selected LLMs under different multi-agent frameworks, with k values on the x-axis and pass@1 on the y-axis.}
    \label{fig_IF}
\end{figure*}

% \begin{figure*}
%     \begin{subfigure}[b]{0.49\textwidth}
%         \includegraphics[width=\linewidth]{image/IF_CodeLlama_7B.pdf}
%         \caption{CodeLlama-Python-7B}
%     \end{subfigure}
%     % \hfill
%     \begin{subfigure}[b]{0.49\textwidth}
%         \includegraphics[width=\linewidth]{image/IF_CodeLlama_13B.pdf}
%         \caption{CodeLlama-Python-13B}
%     \end{subfigure}
    
%     % \vspace{0.5cm}
    
%     \begin{subfigure}[b]{0.49\textwidth}
%         \includegraphics[width=\linewidth]{image/IF_CodeLlama_34B.pdf}
%         \caption{CodeLlama-Python-34B}
%     \end{subfigure}
%     \hfill
%     \begin{subfigure}[b]{0.49\textwidth}
%         \includegraphics[width=\linewidth]{image/IF_DeepSeek_1_3B.pdf}
%         \caption{DeepSeek-Coder-1.3B}
%     \end{subfigure}
    
%     % \vspace{0.5cm}
    
%     \begin{subfigure}[b]{0.49\textwidth}
%         \includegraphics[width=\linewidth]{image/IF_DeepSeek_6_7B.pdf}
%         \caption{DeepSeek-Coder-6.7B}
%     \end{subfigure}
%     % \hfill
%     \begin{subfigure}[b]{0.49\textwidth}
%         \includegraphics[width=\linewidth]{image/IF_DeepSeek_33B.pdf}
%         \caption{DeepSeek-Coder-33B}
%     \end{subfigure}

%     \caption{Line graphs of accuracy rates for each open-source LLM under different multi-agent frameworks, with k values on the x-axis and test accuracy rates on the y-axis.}
%     \label{fig_IF}
% \end{figure*}

The iterative workflow is one of the key designs of all multi-agent frameworks. We investigated the impact of the iteration number and the underline limitations. 

% As mentioned in Section \ref{multi_agent_framework}, the algorithmic complexity of AgentCoder, INTERVENOR, and Self-Collaboration is $O(k)$, while MapCoder has a complexity of $O(kt)$. Here, $k$ represents the number of iterations in the refinement operation, which allows LLMs to iteratively fix problematic code.

% As mentioned in subsection \ref{multi_agent_framework}, all four selected multi-agent frameworks incorporate an iterative refinement process that allows the model to integrate previously failed code and error messages. In our initial experiments, we uniformly set the number of iterative refinements, $k$, to 5 for all four multi-agent frameworks. However, we now suspect that this process may not have achieved its intended effect because of the poor performance in RQ1. To verify this hypothesis, we conducted additional tests by varying $k$ values to 1, 2, 3, and 4, and recorded the results. We plotted line graphs of accuracy rates for each open-source LLM under different multi-agent frameworks, with $k$ values on the x-axis and test accuracy rates on the y-axis, as shown in Figure \ref{fig_IF}. 
\vspace{5pt}\noindent\textbf{Impact on Varied Iterations.}
We assume that this process may not have achieved its intended effect due to the poor performance observed in RQ1. To verify this hypothesis, we performed an in-depth analysis. By varying the number of iterations $k$ from 1 to 5, we recorded the experimental results. We plot line graphs of pass@1 for each selected LLM in different multi-agent frameworks,
% , with $k$ values on the x-axis and test accuracy on the y-axis, 
as shown in Figure \ref{fig_IF}. 
% Analysis of these line graphs 
It reveals that the iterative refinement process has a 
% extremely 
limited impact on improving LLM's code generation capabilities in most cases. For example, Self-Collaboration using CodeLlama-Python-7B 
% as its foundation LLM 
maintains a constant pass@1 of 31.71\%, regardless of changes in $k$. More notably, for MapCoder based on CodeLlama-Python-13B, increasing $k$ leads to a decrease in the pass@1 of code generation. Even for the combination that showed the most significant improvement, AgentCoder with CodeLlama-Python-34B, the pass@1 only increased from 41.46\% to 48.17\% after five rounds of iteration. This result represents a total improvement of 16.18\%, or an average of merely 3.82\% per round. In addition, each additional round of iterative refinement results in a multiplying increase in both token consumption and inference time. 

\vspace{5pt}\noindent\textbf{Limitation of Iterative Refinement.}
% To investigate the reasons behind the limited effectiveness of iterative refinement, 
To further investigate the effectiveness of iterative refinement, 
we conducted a analysis of the detailed response content from each multi-agent framework during code generation tasks. Focusing particularly on samples that failed to be successfully repaired, we extracted pairs of code before and after refinement for in-depth manual analysis. Based on our meticulous examination of 1,859 code pairs, we categorized the model's refinement performance into five types, considering both the code content before and after refinement and their respective test results. This classification is presented in detail in Table \ref{tab_refine}.

\begin{table*}
    \centering
    \small
    \caption{Classification of six selected LLMs' performance in iterative refinement.} 
    \begin{tabular}{lp{8.5cm}cr}
        \toprule
         \textbf{Refinement Type} & \textbf{Description} & \textbf{Proportion} & \textbf{Example} \\
        \midrule
         Code Invariance & Cases where the code remains entirely unchanged before and after refinement & 326 (17.54\%) & Fig. \ref{fig_code_invariance} in Appendix \\
        \midrule
         Error Message Persistence & Excluding the above category, cases where the error messages remain completely identical before and after refinement & 428 (23.02\%) & Fig. \ref{fig_error_message_persistence} in Appendix \\
        \midrule
         Error Type Consistency & Excluding the previous two categories, cases where the error types remain consistent before and after refinement & 281 (15.12\%) & Fig. \ref{fig_error_type_consistency} in Appendix \\
        \midrule
         Function Emptying & Excluding the previous three categories, cases where the function becomes completely empty after refinement & 91 (4.90\%) & Fig. \ref{fig_function_emptying} in Appendix \\
        \midrule
         Miscellaneous Refinement & Other refinement scenarios not covered by the previous four categories & 733 (39.42\%) & Fig. \ref{fig_miscellaneous_refinement} in Appendix \\
        \bottomrule
    \end{tabular}
    \label{tab_refine}
\end{table*}

% Through an in-depth analysis of the descriptions for each Refinement Type in the table, we can infer that two categories represent normal cases of unsuccessful refinement: Error Type Consistency and Miscellaneous Refinement. 
According to Table \ref{tab_refine}, Error Type Consistency and Miscellaneous Refinement are the two most common normal refinement types.
The former indicates that the LLM attempted to fix a previous issue but failed,
% to do so successfully, 
while the latter suggests that the LLM's attempt to address the original problem resulted in the introduction of new types of errors. In contrast, Code Invariance, Error Message Persistence, and Function Emptying are considered abnormal behaviors in the iterative refinement process. Code Invariance refers to instances where the code remains identical before and after refinement; Error Message Persistence indicates that the error message remains unchanged, which, given that error messages contain information about the error type and location, implies minimal changes to the code; Function Emptying describes cases where the refined function implementation is empty, completely failing to achieve the intended refinement effect. Notably, these three abnormal categories collectively account for a substantial 45.46\% of the cases, approaching half of the sample. 
% \zyh{(ZYH: TODO)}
% This high proportion of ineffective refinements largely explains the poor performance of the iterative refinement process.
This high proportion of ineffective refinements further illustrates the ineffectiveness of the process of iterative refinement.

\vspace{5pt}
\subsubsection{Effectiveness Analysis of MapCoder} \label{mapcoder}

Section \ref{accuracy_performance} reveals that MapCoder achieves the most significant improvement on six selected diverse LLMs. 
% The improvement effects of other frameworks are less pronounced, and in some cases, they may even diminish the code generation capabilities of their foundation models, resulting in test accuracies below the baseline. 
It is worthwhile to investigate the factors contributing to MapCoder's effectiveness. 
% will provide valuable insights. for designing our new multi-agent framework.

\begin{figure}
    \centering
    \includegraphics[width=\linewidth]{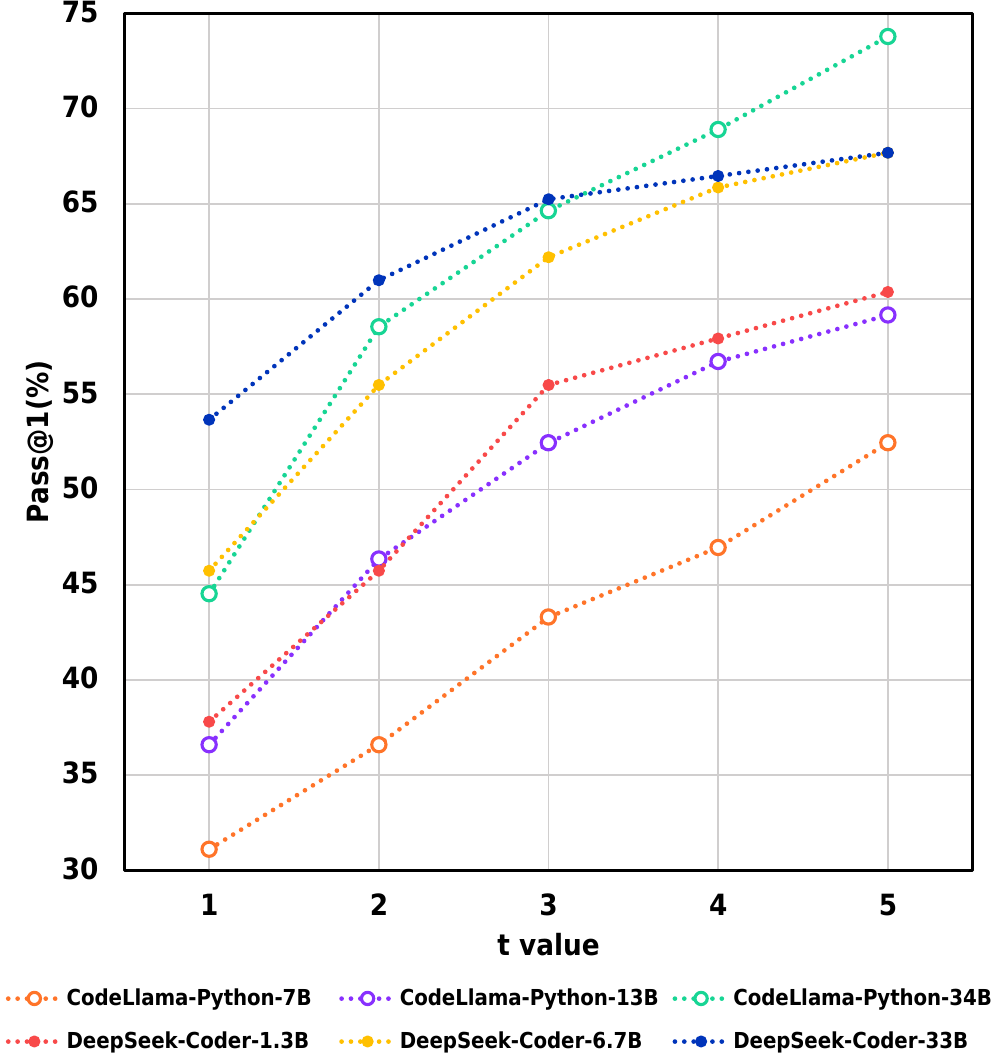}
    \caption{A pass@1 line graph for six selected LLMs under MapCoder's influence, with t values on the x-axis and test accuracies on the y-axis.} % 
    \label{fig_line_mapcoder}
\end{figure}

\vspace{5pt}\noindent\textbf{Impact on Multi-Plan Coding.} %\hy{what is "multi-plan coding", can introduce it in Section 2}
Experimental results presented in Section \ref{iteration_refinement} have confirmed that the iterative refinement process hardly improves the pass@1 of code generation. Instead, it leads to a multiplicative increase in token consumption and inference time. The key distinction between MapCoder and the other three multi-agent frameworks lies in its planning mechanism ``Multi-Plan Coding". This mechanism first instructs an LLM to generate $t$ tasks relevant to the given task prompt, then creates a step-by-step generation plan for each new task, and finally guides the LLM-based code generation. %\hy{what is 'a code'?}

This approach of repeatedly generating new code from the beginning 
% (hereafter referred to as "Multi-Plan Coding" for brevity) 
differs fundamentally from executing refinement operations on existing problem code. %, accounting for the $O(t)$ component in its time complexity $O(kt)$. 
We hypothesize that this innovative operation is the primary factor enabling MapCoder to significantly enhance code generation capabilities. To validate this hypothesis, we conducted additional tests by varying the number of plans from 1 to 5, and recorded experimental results
% . We plotted accuracy line graphs for each open-source LLM under MapCoder's influence, with t values on the x-axis and test accuracies on the y-axis, 
in Figure \ref{fig_line_mapcoder}. 
The analysis of the line graphs demonstrates that Multi-Plan Coding significantly enhances code generation capabilities. Among the models tested, CodeLlama-Python-7B showed the most substantial improvement, with its pass@1 increasing from 31.10\% to 52.44\%, representing a remarkable 68.62\% improvement. Even the DeepSeek-Coder-33B model, which exhibited the smallest improvement, still achieved a considerable 26.13\% increase. 
% Consequently, we can explain MapCoder's significant performance enhancement as follows: By establishing diverse problem-solving plans for the LLM in advance, MapCoder enables the LLM to generate a wider variety of solutions. Subsequently, through an optimization selection mechanism, it ultimately achieves a substantial increase in accuracy.

\begin{figure}
    \centering
    \includegraphics[width=\linewidth]{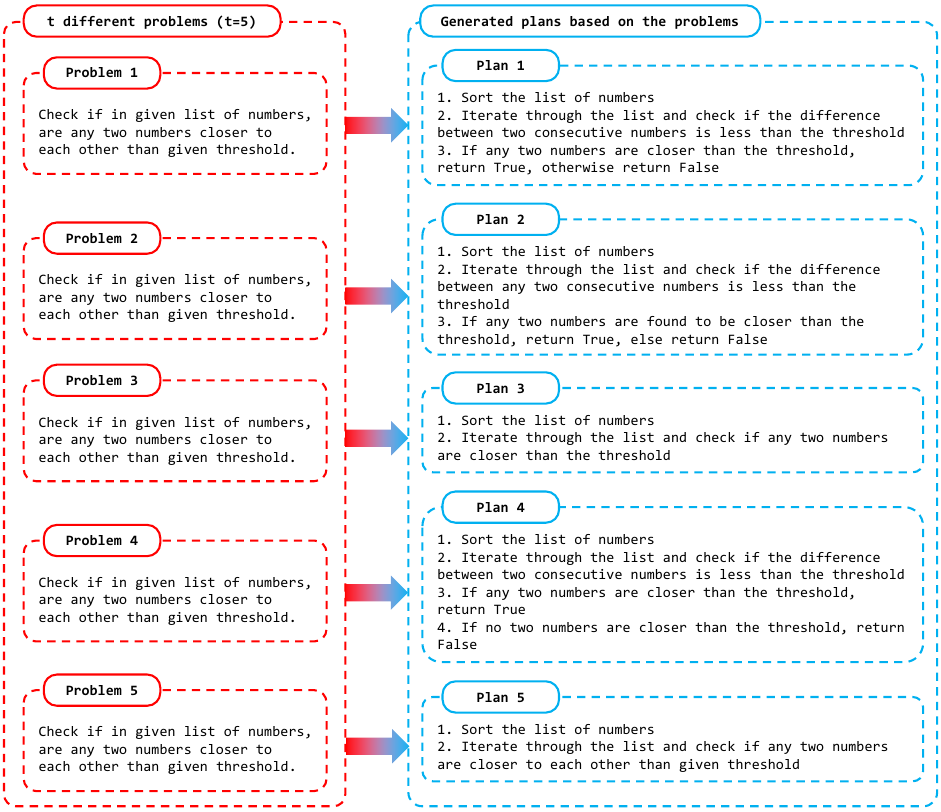}
    \caption{The $t$ different problems and the corresponding plans generated by MapCoder are highly similar.} 
    \label{fig_mapcoder_plans}
\end{figure}

\vspace{5pt}\noindent\textbf{Pros and Cons of Multi-Plan Coding.}
Although MapCoder achieves the best performance, it still has some drawbacks. Firstly, the inclusion of the additional Multi-Plan Coding step in MapCoder increases its time complexity from $O(k)$, as seen in the other three multi-agent frameworks, to $O(kt)$. This significantly increases the number of tokens consumed and the inference time. Secondly, its plan generation mechanism is unreliable because the quality of the plans is scored by another LLM, making the selection of the best plan heavily dependent on the inherent capabilities of the LLMs. Thirdly, the generated plans lack diversity. As shown in Figure \ref{fig_mapcoder_plans}, the $t$ different problems generated by its retrieval agent are highly similar, leading to the generation of identical plans. We assume that different plans could lead to better performance and validate this in Section \ref{evaluation}.

% 2) MapCoder's ``Multi-Plan Coding'' mechanism, which first instructs LLMs to generate different tasks, then develop a plan for each task, and finally produce code based on the best plan as judged by LLMs, has proven to be effective. However, its plan generation mechanism is unreliable because selecting the best plan relies heavily on the inherent capabilities of the LLMs, and the multiple tasks initially generated are often only slightly different, leading to the generation of identical plans;

\begin{figure}
    \centering
    \includegraphics[width=\linewidth]{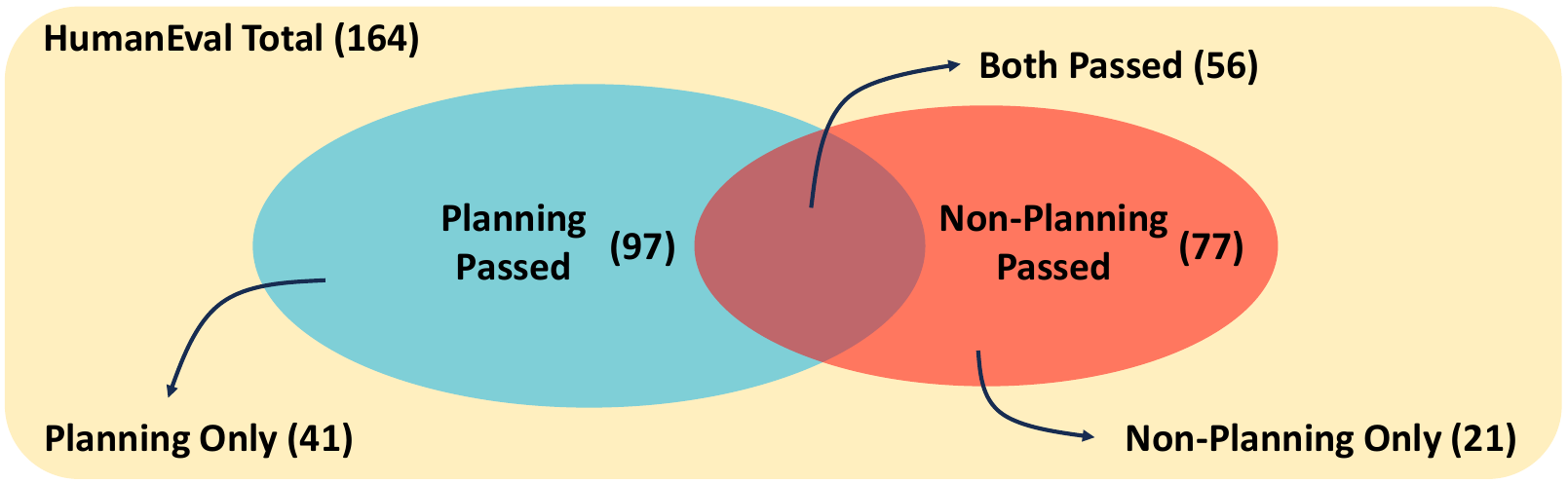}
    \caption{Venn diagram of HumanEval sample pass for Planning and Non-Planning Frameworks.} 
    \label{fig_Venn}
\end{figure}

In addition, we collect the output results for each problem
% in HumanEval 
from the planning framework of MapCoder and the other three non-planning frameworks, based on six selected open-source LLMs. 
For each problem, we extracted one result from each type of framework and counted whether they passed the test or not. 
This process was repeated three times and the average was taken to create the final Venn diagram as shown in Figure \ref{fig_Venn}. We can observe that the number of samples passed by the non-planning frameworks is lower than that of the planning framework (77 vs. 97). However, they are not a subset of each other with an intersection of 56 samples. Thus, 41 and 21 samples exclusively passed for the planning and non-planning frameworks, respectively. This result implies that combining planning and non-planning mechanisms is expected to lead to better performance, with computational resource requirements lower than those of using only the planning mechanism.

\vspace{0.3cm}
\begin{mdframed}[nobreak=true]
\textbf{Answer to RQ2:} Iterative refinement offers no substantial effectiveness but causes significant increase in inference cost. The multi-plan coding provides major contribution to MapCoder but incurs high computational cost; the planing and non-planning generation is complementary to each other.
% noise

%However, this step is unreliable, as the generated plans lack diversity. 
% Selective use of multi-plan coding is cost-effective to also involve a high token consumption and inference time.
\end{mdframed}

% \section{Empirical Findings}

\section{Methodology} \label{method}
% \subsection{AdaCoder} \label{AdaCoder}

\begin{figure*}
    \centering
    \includegraphics[width=\linewidth]{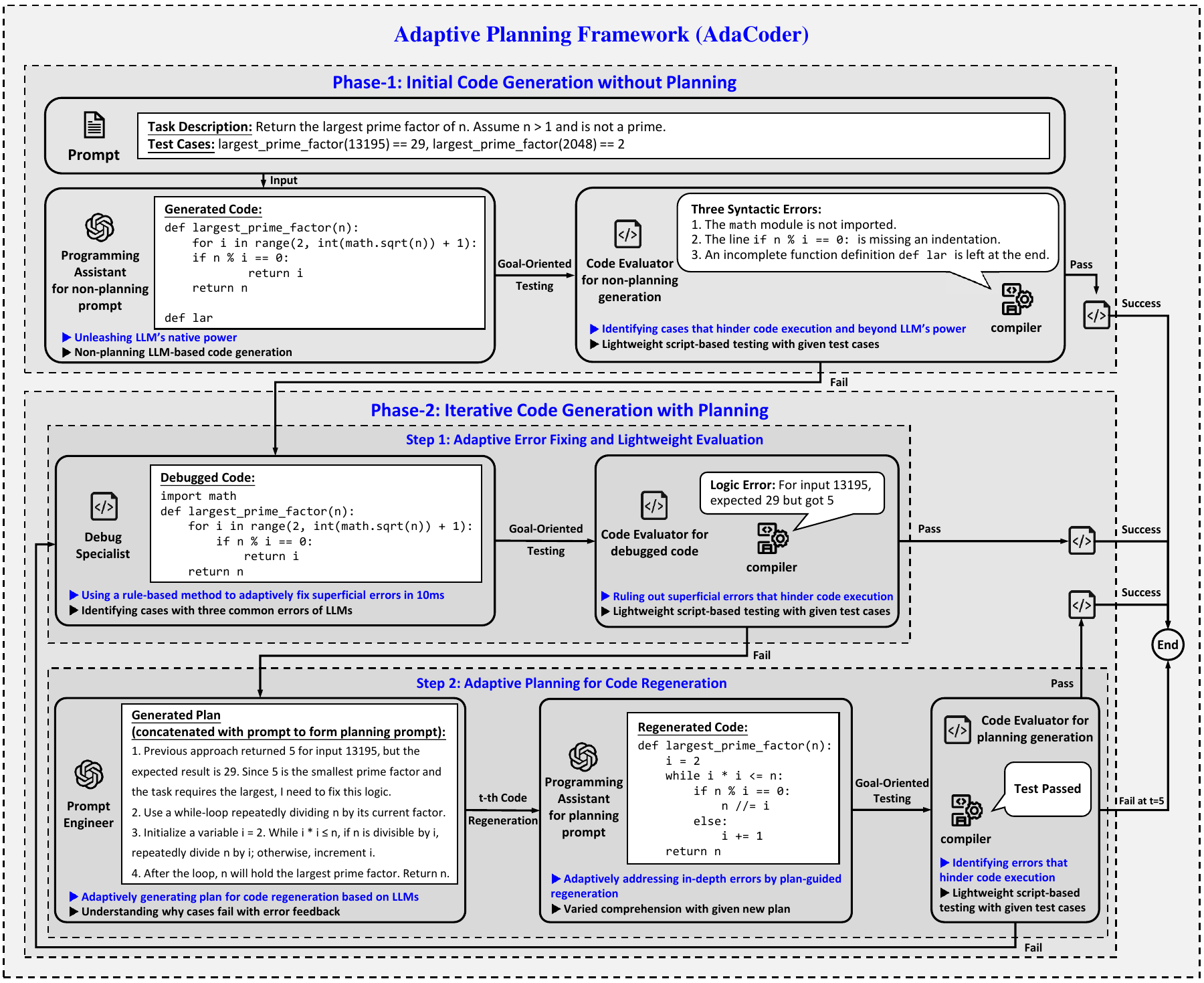}
    \caption{The Workflow of AdaCoder.}%\hy{what does the Yellow texts mean? Some concepts needs to be explained, such as "Goal-Oriented Testing" (if needed at all), etc}} % \lc{it is difficult to read this workflow. please highlight the four agents. give step numbers; clarify the components of the iterative process; how the debug specialist (I cannot see the rule repository related concepts) know the compilation and runtime errors? (this info is collected by the Code Evaluator?) anyway, this overflow should be easy to read without knowing the context and its logic is strictly described in the paragraphs.}
%    \hy{still, this figure shows that the proposed method and related methods are quite similar. The adaptive planning mechanism could be highlighted}
    \label{fig_workflow} 
\end{figure*}

Based on our empirical findings, we design an adaptive planning framework for multi-agent code generation called \textbf{AdaCoder}. Our goal is to develop a cost-effective multi-agent framework that achieves better generalizability across LLMs with varying parameter sizes, architectures, and generation capabilities.
% RQ2 shows that all frameworks leverage iterative refinement, while MapCoder additionally employs planning. Iterative refinement is both ineffective and costly, while planning is effective but may achieve better results when combined with non-planning in an adaptive manner to selectively apply planning. Inspired by these findings, we propose an adaptive planning multi-agent framework, named AdaCoder, which is simpler, faster, more effective and universal than existing frameworks.
% The research findings from RQ2 provided significant insights for designing our new multi-agent framework, AdaCoder, an adaptive planning framework for multi-agent code generation. AdaCoder enhances the plan diversity and leverages the complementarity of planning and non-planning mechanisms. 
%In this study, the foundation LLM is implemented using HuggingFace's model parameters and interfaces for open-source models. For closed-source models such as ChatGPT series LLMs, we utilize the official OpenAI APIs.

\subsection{Overall Workflow}

AdaCoder consists of four collaborative agents to generate code: a Programming Assistant, a Code Evaluator, a Debug Specialist, and a Prompt Engineer. The details can be found in Algorithm \ref{algo_workflow} in Appendix. Figure \ref{fig_workflow} illustrates the overall workflow of AdaCoder using an example task: finding the largest prime factor of a number $n$. The process includes two phases.

Phase-1 focuses on Initial Code Generation without Planning, aiming to leverage the LLM's native capabilities directly. Initially, the Programming Assistant receives the task description and sample test cases. It then generates the initial code without a plan, as depicted in the top-left of Figure \ref{fig_workflow}. This code may contain superficial errors, which are defined as errors in syntax or structure (like missing imports, incorrect indentation, or incomplete function definitions as shown in the example) that prevent the code from being compiled or run correctly. Consequently, these errors hinder code execution, meaning the program cannot produce an output that can be compared against expected test results. Subsequently, the Code Evaluator performs ``Goal-Oriented Testing." In this phase, its primary goal is specifically to detect these execution-hindering superficial errors within try-except blocks. If no such errors are found and the code passes all tests, the process ends. However, the flawed code and specific error details are passed forward to initiate Phase-2.

Phase-2 involves Iterative Code Generation with Planning and is triggered only upon Phase-1 failure. It consists of two steps, repeated up to t times. First, in Step 1, the Debug Specialist takes the code and error information, applying rule-based fixes for common superficial issues (e.g., adding import math, correcting indentation) to produce syntactically correct code. The Code Evaluator then performs ``Goal-Oriented Testing" again, but now with the goal shifted to detecting logic errors (also referred to as in-depth errors). These are defined as flaws in the algorithm's reasoning or implementation that cause the code to produce incorrect results (i.e., fail test case assertions), even if it runs without crashing. For instance, the debugged code in Figure \ref{fig_workflow} runs but returns 5 instead of the expected 29, indicating a logic error. If the tests pass, the process ends. If logic errors are detected, the failure feedback proceeds to the next step. In Step 2, the Prompt Engineer uses the original task description and the specific logic error feedback to generate a tailored step-by-step plan to correct the issue (like the while-loop plan in Fig. \ref{fig_workflow}). The Programming Assistant then regenerates the code, guided by this plan. Finally, the Code Evaluator tests this new code. Success leads to termination; failure leads back to the beginning of Phase-2 (Step 1) with the latest error information, iterating until success or the maximum t attempts are reached.
% The collaboration process has two phases. As shown in Fig. \ref{fig_workflow}, Phase-1 is an initial code generation without planning, using an LLM-based agent \textit{Programming Assistant} to unleash the LLM's native power by generating code using a prompt without plan, and a script-based agent \textit{Code Evaluator} to identify cases beyond LLM's power and errors hindering execution. To reduce inference costs and ensure accurate information transfer, the Code Evaluator assesses code correctness using the sample test cases given in the prompt, instead of LLM-based test case generation as AgentCoder. The information of detected errors are then feedback to the Debug Specialist in Phase-2. Phase-2 iteratively addresses superficial and in-depth errors by adaptive error fixing and adaptive planning. The \textit{Debug Specialist} adaptively fixes three superficial errors using a rule-based method derived from our prior work \cite{wen2024fixingcodegenerationerrors}, based on the error feedback. This replaces costly LLM-based bug localization \cite{qin2024agentfl} and debugging \cite{lee2024unified}. Subsequently, the \textit{Prompt Engineer} generates a step-by-step plan to guide the Programming Assistant to address in-depth errors. This plan explains the logical flaws (returning the smallest prime factor instead of the largest) and outlines correct steps to resolve them, enabling the \textit{Programming Assistant} to regenerate code based on this plan and ultimately pass all tests.

Generally, AdaCoder employs ``Adaptive Planning", which is achieved through two strategies: 1) It only applies the planning mechanism for iterative regeneration when the LLM's native capability proves insufficient (i.e., the initial non-planning generation fails), rather than using planning for every attempt; 2) During the planning phase, it generates a plan that is adapted to the specific error feedback from the failed regeneration attempt.

\subsection{Programming Assistant}

The Programming Assistant is an LLM-based agent responsible for generating code. It takes the task description provided by the benchmark (such as HumanEval) and the plan formulated by the Prompt Engineer (if any) as input and outputs a code corresponding to the task.

\vspace{5pt}\noindent\textbf{Technical Implementation.} This agent is powered by an arbitrary LLM, which generates code based on a given prompt, as illustrated in Fig. \ref{fig_PA_prompt} in Appendix. The prompt can take two forms: 1) When generating code for a given task for the first time, the prompt consists solely of the task description. This approach corresponds to the non-planning mechanism, as illustrated in Fig. \ref{fig_PA_prompt}(a) in Appendix. 2) When the initial code generation fails, even after the Debug Specialist has been applied, the prompt is formed by concatenating the task description with the step-by-step plan devised by another LLM-based agent, the Prompt Engineer. This approach corresponds to the planning mechanism, as shown in Fig. \ref{fig_PA_prompt}(b) in Appendix.

\vspace{5pt}\noindent\textbf{Design Rationale.} The Programming Assistant employs the non-planning mechanism during the initial code generation and adaptively switches to the planning mechanism for subsequent regeneration attempts. This design aligns with the findings in RQ2 that ``combining planning and non-planning mechanisms shows potential for better performance and reduced computational resources", effectively leveraging the strengths of both approaches.

\subsection{Code Evaluator}

The Code Evaluator is a script-based agent designed to assess the correctness of the code. It takes the code generated by the Programming Assistant (or the code debugged by the Debug Specialist) and the sample test cases provided in the benchmark as input, and outputs the test results (pass/fail) along with the error information (if the test fails).

\vspace{5pt}\noindent\textbf{Technical Implementation.} The pseudocode for this agent is presented in Algorithm \ref{alg_code_evaluator} in the appendix. It first embeds the code generated by the Programming Assistant (or debugged by the Debug Specialist) into a try-except block for compilation to collect error feedback. If no exception is captured, it indicates a successful compilation. The sample test cases provided by the benchmark (e.g. HumanEval) are then appended to the code and executed within another try-except block. If no exceptions are raised during execution, it indicates that the code passes the test. Otherwise, the try-except block captures detailed compilation or runtime error information, such as ``SyntaxError: unterminated triple-quoted string literal (detected at line 68) ($<$string$>$, line 41)". If the tested code is provided by the Programming Assistant, this information is fed back to the Debug Specialist for debugging. If the tested code is provided by the Debug Specialist, the information is instead fed back to the Prompt Engineer for planning.

\vspace{5pt}\noindent\textbf{Design Rationale.} Existing research (e.g., AgentCoder \cite{huang2024agentcodermultiagentbasedcodegeneration}) often relies on LLMs to automatically generate test cases. However, test case generation is as challenging as code generation \cite{pacheco2007feedback}, and incorrect test cases can lead to erroneous code \cite{islam2024mapcodermultiagentcodegeneration}. Blindly editing code based on these test cases can undermine problem-solving capabilities \cite{shinn2024reflexion}. Therefore, we use benchmark-provided sample test cases to evaluate code correctness, consistent with the approach of other frameworks (e.g., INTERVENOR \cite{wang2024intervenorpromptingcodingability}, MapCoder \cite{islam2024mapcodermultiagentcodegeneration}, Self-Collaboration \cite{dong2024self}). 
These test cases are extracted from the \texttt{prompt} field, thereby avoiding the risk of the LLM generating incorrect test cases that could introduce noise.
Additionally, while existing methods \cite{islam2024mapcodermultiagentcodegeneration, dong2024self} merely evaluate the code, the Code Evaluator incorporates a try-except block to automatically collect the error information for debugging or planning, making full use of the available test information.

\subsection{Debug Specialist}

The Debug Specialist is a rule-based agent used to fix simple errors in the code. It takes the code generated by the Programming Assistant and the error information collected by the Code Evaluator as input, and outputs the debugged code.

\vspace{5pt}\noindent\textbf{Technical Implementation.} This agent is a script derived from our previous research, LlmFix \cite{wen2024fixingcodegenerationerrors}, which addresses three types of common and simple code errors: Inconsistent Indentation, Function Overflow, and Missing Import. % \hy{are they common errors? why only these three errors? It seems that all errors discussed in this section are syntax-related errors?}. 
Our previous studies \cite{wen2024fixingcodegenerationerrors} show that only these three types of syntax-related errors are well-suited for lightweight, rule-based fixes. For other more complex logic errors, we employ the planning agent Prompt Engineer, as discussed in Section \ref{PE}.
The Debug Specialist resolves the three types of errors through the following three steps.
% Specifically, it implements a three-stage error correction pipeline for the code generated by the Programming Assistant: 
% 1) \textbf{Code Filtering} standardizes indentation patterns (tabs/spaces conversion) and removes extraneous constructs (print/assert statements) to resolve IndentationErrors and execution blocking; 
\textit{1) Code Filtering}. %This step splits the code line by line, storing it in a list, and checks the indentation of each line. Typically, the indentation should be four spaces. If a line's indentation does not conform to this standard, 
The Debug Specialist checks and normalizes the indentation according to the logic outlined in Algorithm \ref{alg_debug} in Appendix, addressing Inconsistent Indentation errors.
% 2) \textbf{Code Truncation} employs an iterative function pruning strategy that progressively removes trailing incomplete functions until successful compilation or single-function retention, addressing SyntaxErrors from function overflow; 
\textit{2) Code Truncation}. Based on our prior research \cite{wen2024fixingcodegenerationerrors}, syntax errors often occur when LLMs exceed output length limits, leading to truncated functions with incomplete syntax at the end of the code, namely Function Overflow. To address this, we need to remove incomplete functions from the end of the code. However, such incomplete functions can take various forms (e.g., different function names and truncation patterns), making it difficult to extract them using only regular expressions. %Instead, 
We achieve this by iteratively removing the last line of the code and compiling the modified code to check if the incomplete function has been fully removed: if the code compiles successfully, it indicates that no incomplete functions remain, as they would cause syntax issues and prevent compilation. If the code still fails to compile, it suggests that incomplete functions remain at the end and further removal is needed. Our prior research has shown that this removal process is fast, with an average execution time of approximately 10ms. %\hy{what is Functional Overflow?} \hy{why removing the last line?}
% Based on our prior research \cite{wen2024fixingcodegenerationerrors}, syntax errors often arise from Function Overflow, where LLMs exceed output length limits, resulting in truncated functions with incomplete syntax at the end of the code. To address this, this step first attempts to compile the code. If the compilation fails, it splits the code into lines, stores them in a list, and removes the last line. This compile-remove loop continues until the code compiles successfully or only one function remains.
% 3) \textbf{Missing Modules Injection} dynamically inserts import statements for Python standard library components (identified via a pre-built database) when NameErrors occur. The pipeline combines static code transformation with error-feedback-driven repairs to maximize code validity. Eventually, it feeds back to the next agent Prompt Engineer the error information (if any) it gets from testing the debugged code. 
\textit{3) Missing Modules Injection}. If the error type provided by the Code Evaluator is a NameError, it indicates that the code may be using a module or function that has not been imported (i.e., Missing Import). This step extracts the name causing the NameError (e.g., math, re, functools) through the regular expression ``\texttt{name '(.+?) ' is not defined}" and attempts to match it against a pre-built database containing all common module names and their internal function names. If the match is found, the corresponding import statements (e.g., \texttt{import math}) is inserted. Otherwise, it indicates that the name refers to an undefined variable rather than a module or library function.
The pseudocode of the Debug Specialist’s workflow is presented in Algorithm \ref{alg_debug} in Appendix. %\lc{the process is difficult to understand.}

\vspace{5pt}\noindent\textbf{Design Rationale.} The Debug Specialist employs a rule-based method to fix errors in generated code. According to the empirical findings in RQ2, using LLMs to iteratively fix errors (i.e., iterative refinement) is ineffective and costly, achieving only an average improvement of 1.6\% over five iterations. In contrast, our prior research \cite{wen2024fixingcodegenerationerrors} demonstrates that a rule-based fixing method yields a 7.5\% improvement while requiring only 11.50 ms per fix. This highlights that the code generation capabilities of low-performance LLMs are unstable and unreliable, resulting in poor performance during iterative refinement. In comparison, rule-based methods are more deterministic, interpretable, and efficient. 

\subsection{Prompt Engineer} \label{PE}

The Prompt Engineer is an LLM-based agent responsible for creating step-by-step plans to complete tasks. It takes the task description from the benchmark and the error information collected by the Code Evaluator after testing the debugged code as input, and outputs a step-by-step plan to accomplish the task.

\vspace{5pt}\noindent\textbf{Technical Implementation.} This agent is powered by an arbitrary LLM consistent with the Programming Assistant. The agent generates a step-by-step plan based on a given prompt, as illustrated in Fig. \ref{fig_PE_prompt} in Appendix. The prompt is formed by concatenating the task description with the error information fed by the Code Evaluator.

\vspace{5pt}\noindent\textbf{Design Rationale.} This plan generation mechanism differs from existing mechanisms, such as the one used in MapCoder. MapCoder first generates multiple similar tasks based on the original task description and then creates a plan for each of these tasks. However, according to the empirical findings in RQ2, the similar tasks generated by MapCoder are often identical, leading to nearly identical plans. As a result, its Coding Agent does not significantly improve its understanding of the task, and the generated code remains largely unchanged. In contrast, AdaCoder's planning mechanism is based on explicit actual error feedback. Since the errors encountered in the code vary each time, this enhances the diversity of the generated plans, enabling the Programming Assistant to attempt different plans during the iterative regeneration process and ultimately improving generation performance.

\section{Evaluation} \label{evaluation}

\subsection{Research Questions}

% This section aims to investigate the following RQs.

% \vspace{0.2cm}
% \begin{mdframed}[nobreak=true]
% \textbf{RQ3: Can AdaCoder outperform existing multi-agent frameworks on open-source and closed-source LLMs?}
% \end{mdframed}

\textbf{RQ3: Can AdaCoder outperform existing multi-agent frameworks on diverse foundation LLMs?} Our empirical studies show that existing multi-agent frameworks often design a large number of agents centered around LLMs, incorporate complex workflow (e.g., iterative generation of multiple results \cite{islam2024mapcodermultiagentcodegeneration} and repeated self-refinement processes \cite{huang2024agentcodermultiagentbasedcodegeneration, islam2024mapcodermultiagentcodegeneration, wang2024intervenorpromptingcodingability, dong2024self}). This significantly increases the time and GPU resources required for code generation \cite{islam2024mapcodermultiagentcodegeneration}. Simpler and faster approaches are generally more practical for real-world production applications. Thus, we present AdaCoder in Section \ref{method} according to our empirical findings. This RQ aims to investigate the effectiveness and cost of AdaCoder and confirm the assumptions behind the design.

% \vspace{0.2cm}
% \begin{mdframed}[nobreak=true]
% \textbf{RQ4: Are all agents of AdaCoder necessarily required?}
% \end{mdframed}

% Section \ref{method} indicates that Programming Assistant and Code Evaluator are nece

\textbf{RQ4: Are all agents of AdaCoder necessary?} As described in Section \ref{method}, among AdaCoder's four agents, the Programming Assistant and Code Evaluator are the core components thus cannot be removed. In contrast, the Prompt Engineer and Debug Specialist are designed to enhance code generation capabilities and can be removed in ablation studies. Consequently, we selected a total of ten LLMs from RQ1 and RQ3 as foundation models to evaluate the pass@1 of AdaCoder under three conditions on the HumanEval dataset: without the Prompt Engineer, without the Debug Specialist, and without both the Prompt Engineer and Debug Specialist. 

%Fu et al. \cite{fu2017easy} also emphasized that in software engineering tasks, it is essential to first explore simple methods to solve problems. %Therefore, we want to investigate: 

\subsection{Experimental Settings}

We apply AdaCoder and the other four multi-agent frameworks to six diverse LLMs as adopted in Section \ref{exp} and four ChatGPT series LLMs including GPT-3.5-turbo, GPT-4, GPT-4-turbo, and GPT-4o. We evaluated the performance of these frameworks on HumanEval and another dataset MBPP. %We regard MapCoder as our baseline multi-agent framework due to its best performance on open- and closed-source LLMs.
% on HumanEval and MBPP dataset to test their performance. This approach aims to demonstrate AdaCoder's effectiveness and generalization capability. 
The MBPP dataset is also widely used \cite{brown2020language,chen2021evaluatinglargelanguagemodels,chowdhery2022palmscalinglanguagemodeling}. It is a comprehensive collection of 974 Python programming tasks designed to evaluate code generation and program synthesis capabilities \cite{mbpp}. %Each task in the dataset is accompanied by an average of three test cases, ensuring the functional correctness of the generated code.

% Prominent language models such as GPT-3 \cite{brown2020language}, Codex \cite{chen2021evaluatinglargelanguagemodels}, and PaLM \cite{chowdhery2022palmscalinglanguagemodeling} have utilized MBPP for assessing their code generation proficiency. 

% The MBPP dataset comprises six essential components: \textit{1) task\_id}, a unique identifier assigned to each task, facilitating easy reference and organization of the code generation challenges. \textit{2) text}, a paragraph of text that provides generation requirements for guiding LLMs. \textit{3) code}, the standard solution for each task. \textit{4) test\_list}, a testing list including several test cases. \textit{5) test\_setup\_code}, a piece of code that runs before executing the tests, ensuring that the tests are executed in the correct context. \textit{6) challenge\_test\_list}, a set of additional, more challenging test cases used to deeply evaluate the robustness and correctness of the generated code.

\subsection{Performance of AdaCoder on Ten Diverse LLMs (RQ3)}\label{eva_accuracy}

\subsubsection{Pass@1 Analysis of Code Generation}

\begin{table}
    % \scriptsize
    \centering
    \caption{The pass@1 performance of AdaCoder compared to the direct method when applied to ten selected diverse LLMs.}  

    \fontsize{6}{7}\selectfont
    \begin{tabular}{l|c|c|c|c}
        \toprule
        \multirow{2}{*}{\textbf{LLMs}} & \multicolumn{2}{c|}{\textbf{Direct}} & \multicolumn{2}{|c}{\textbf{AdaCoder}} \\ 
        \cmidrule{2-5} & HumanEval & MBPP & HumanEval & MBPP \\
        \midrule
        CodeLlama-Python-7B  & 32.69 & 42.12 & 63.41 (\textcolor{cgreen}{$\uparrow$93.97\%}) & 68.40 (\textcolor{cgreen}{$\uparrow$62.39\%}) \\
        CodeLlama-Python-13B & 36.65 & 46.86 & 71.95 (\textcolor{cgreen}{$\uparrow$96.32\%}) & 70.40 (\textcolor{cgreen}{$\uparrow$50.23\%}) \\
        CodeLlama-Python-34B & 43.72 & 49.82 & 81.10 (\textcolor{cgreen}{$\uparrow$85.50\%}) & 76.40 (\textcolor{cgreen}{$\uparrow$53.35\%}) \\
        DeepSeek-Coder-1.3B  & 49.39 & 47.80 & 76.83 (\textcolor{cgreen}{$\uparrow$55.56\%}) & 71.80 (\textcolor{cgreen}{$\uparrow$50.21\%}) \\
        DeepSeek-Coder-6.7B  & 58.54 & 58.60 & 85.98 (\textcolor{cgreen}{$\uparrow$46.87\%}) & 78.00 (\textcolor{cgreen}{$\uparrow$33.11\%}) \\
        DeepSeek-Coder-33B   & 64.63 & 66.40 & 90.85 (\textcolor{cgreen}{$\uparrow$40.57\%}) & 81.40 (\textcolor{cgreen}{$\uparrow$22.59\%}) \\
        % $\Delta$ Open-Source Avg.    & - & - & \textcolor{cgreen}{$\uparrow$69.80\%} & \textcolor{cgreen}{$\uparrow$45.31\%} \\
        % \midrule
        GPT-3.5-turbo        & 60.30 & 52.20 & 96.95 (\textcolor{cgreen}{$\uparrow$60.78\%}) & 89.40 (\textcolor{cgreen}{$\uparrow$71.26\%}) \\
        GPT-4                & 67.00 & 68.30 & 96.95 (\textcolor{cgreen}{$\uparrow$44.70\%}) & 90.40 (\textcolor{cgreen}{$\uparrow$32.36\%}) \\
        GPT-4-turbo          & 87.10 & 63.40 & 98.17 (\textcolor{cgreen}{$\uparrow$12.71\%}) & 90.40 (\textcolor{cgreen}{$\uparrow$42.59\%}) \\
        GPT-4o               & 90.20 & 67.20 & 98.17 (\textcolor{cgreen}{$\uparrow$08.84\%}) & 91.40 (\textcolor{cgreen}{$\uparrow$36.01\%}) \\
        % $\Delta$ Closed-Source Avg.    & - & - & \textcolor{cgreen}{$\uparrow$31.76\%} & \textcolor{cgreen}{$\uparrow$45.56\%} \\
        \midrule
        $\Delta$ Average    & - & - & \textcolor{cgreen}{$\uparrow$54.58\%} & \textcolor{cgreen}{$\uparrow$45.41\%} \\
        \bottomrule
    \end{tabular}
\label{tab_AdaCoder_accuracy}
\end{table}

\begin{table*}
    \scriptsize
    \centering
    \caption{The pass@1 performance of selectd four multi-agent frameworks compared to the direct method when applied to the six selected open-source models on the MBPP benchmark.}
    \begin{tabular}{lccccc}
        \toprule
        \textbf{LLMs} & \textbf{Direct} & \textbf{AgentCoder} & \textbf{MapCoder} & \textbf{INTERVENOR} & \textbf{Self-Collaboration} \\
        \midrule
        CodeLlama-Python-7B & 42.12 & 36.80 (\textcolor{red}{$\downarrow$12.63\%}) & 55.00 (\textcolor{cgreen}{$\uparrow$30.58\%}) & 47.00 (\textcolor{cgreen}{$\uparrow$11.59\%}) & 35.20 (\textcolor{red}{$\downarrow$16.43\%}) \\
        CodeLlama-Python-13B & 46.86 & 36.60 (\textcolor{red}{$\downarrow$21.90\%}) & 61.60 (\textcolor{cgreen}{$\uparrow$31.46\%}) & 53.00 (\textcolor{cgreen}{$\uparrow$13.10\%}) & 40.40 (\textcolor{red}{$\downarrow$13.79\%}) \\
        CodeLlama-Python-34B & 49.82 & 51.60 (\textcolor{cgreen}{$\uparrow$03.57\%}) & 66.40 (\textcolor{cgreen}{$\uparrow$33.28\%}) & 56.40 (\textcolor{cgreen}{$\uparrow$13.21\%}) & 43.20 (\textcolor{red}{$\downarrow$13.29\%}) \\
        DeepSeek-Coder-1.3B & 47.80 & 21.60 (\textcolor{red}{$\downarrow$54.81\%}) & 37.40 (\textcolor{red}{$\downarrow$21.76\%}) & 37.00 (\textcolor{red}{$\downarrow$22.59\%}) & 38.60 (\textcolor{red}{$\downarrow$19.25\%}) \\
        DeepSeek-Coder-6.7B & 58.60 & 60.20 (\textcolor{cgreen}{$\uparrow$02.73\%}) & 44.60 (\textcolor{red}{$\downarrow$23.89\%}) & 67.80 (\textcolor{cgreen}{$\uparrow$15.70\%}) & 53.40 (\textcolor{red}{$\downarrow$08.87\%}) \\
        DeepSeek-Coder-33B & 66.40 & 48.00 (\textcolor{red}{$\downarrow$27.71\%}) & 60.40 (\textcolor{red}{$\downarrow$09.04\%}) & 61.20 (\textcolor{red}{$\downarrow$07.83\%}) & 53.20 (\textcolor{red}{$\downarrow$19.88\%}) \\
        \midrule
        $\Delta$ Average & - & \textcolor{red}{$\downarrow$18.46\%} & \textcolor{cgreen}{$\uparrow$06.77\%} & \textcolor{cgreen}{$\uparrow$03.86\%} & \textcolor{red}{$\downarrow$15.25\%} \\
        \bottomrule
    \end{tabular}
\label{tab_other_mbpp_accuracy}
\end{table*}

Table \ref{tab_AdaCoder_accuracy} presents the results of AdaCoder with a maximum of $t = 5$ iterations on the HumanEval and MBPP datasets. The performance of the baseline multi-agent frameworks on the two datasets is shown in Tables \ref{tab_overview_a} and \ref{tab_other_mbpp_accuracy}, respectively. 
We can observe that AdaCoder improves the performance of ten LLMs by 54.58\% on HumanEval and 45.41\% on MBPP, totaling an average improvement of 50.00\%.
In contrast, MapCoder, the most effective baseline framework, only achieves an average improvement of 50.43\% on HumanEval and 14.04\% on MBPP, resulting in a total average improvement of 32.24\%. These results demonstrate that AdaCoder outperforms the state-of-the-art multi-agent frameworks.

\vspace{5pt}
\subsubsection{Impact on Iteration Number $t$}

\begin{figure}
    \centering
    \includegraphics[width=\linewidth]{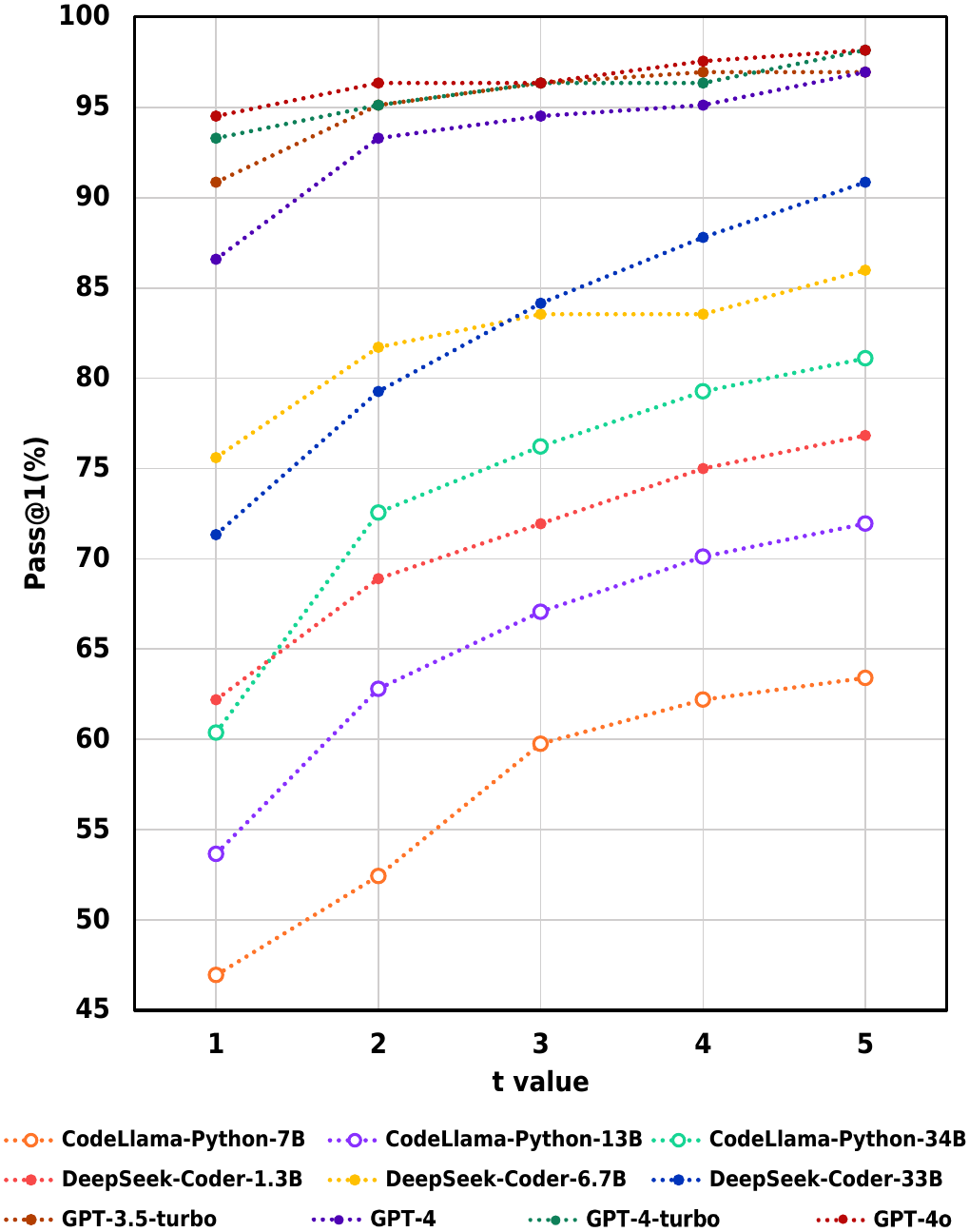}
    \caption{Line graphs with $t$ values as the x-axis and pass@1 as the y-axis.}
    \label{fig_AdaCoder_t_value}
\end{figure}

In RQ2, we validated the ineffectiveness of Iterative Refinement by conducting repeated experiments with varying iteration counts $k$ across four selected multi-agent frameworks. Similarly, we performed repeated tests on the HumanEval dataset by adjusting the maximum iterations $t$ from 1 to 5, 
as shown in Figure \ref{fig_AdaCoder_t_value}.

We observe that for ChatGPT series LLMs, increasing values of $t$ has a minimal impact. For example, when AdaCoder uses GPT-4o as the foundation model, increasing $t$ from 1 to 5 only improves pass@1 from 94.51\% to 98.17\%, a mere 3.87\% increase. Among the four closed-source LLMs, only GPT-4 showed a notable improvement when $t$ increased from 1 to 2: from 86.59\% to 93.29\% with a 7.74\% increase. Statistical analysis indicates that each increase in the $t$ value results in an average improvement of only 1.69\% in code generation pass@1. However, it is noteworthy that even with a single execution of the workflow shown in Figure \ref{fig_workflow}, these closed-source models perform exceptionally well, far surpassing baseline results. For example, GPT-4-turbo and GPT-4o achieved 93.29\% and 94.51\% precision, respectively, with $t$ set to 1. Therefore, these results imply that the limited improvement effect of AdaCoder on closed-source models is due to their inherently strong code generation capabilities. After just one round of the workflow shown in Figure \ref{fig_workflow}, their pass@1 is already very high, leaving limited room for improvement and resulting in the less pronounced enhancement effect of AdaCoder. However, for the other six LLMs, increasing the values of $t$ significantly enhances the code generation capabilities. For example, when AdaCoder uses CodeLlama-Python-34B as the foundation LLM,  increasing $t$ from 1 to 2 improves pass@1 from 60.37\% to 72.56\%, an approximately 20\% increase. Even for the DeepSeek-Coder-6.7B, which showed the smallest improvement, increasing $t$ from 1 to 5 still yielded a 13.72\% performance boost. Statistical analysis indicates that each increase in the $t$ value results in an average 6. 33\% improvement in the pass@1 of the generation. Furthermore, we compared the improvement magnitudes when increasing $t$ from 1 to 3 and from 3 to 5, finding them to be 20.44\% and 6.24\% respectively, with the latter significantly lower. This shows a marginal effect on the improvement of code generation capabilities as the values of $t$ increase.

\vspace{5pt}
\subsubsection{Cost Analysis of Code Generation}\label{eva_resource}

% Although AdaCoder significantly enhances the performance of both open-source and closed-source LLMs, its practical value could be questioned if it consumes excessive computational resources and inference time similar to MapCoder. Therefore, 
To assess the inference cost of AdaCoder, we recorded the average token consumption and inference time. As the inference time of closed-source LLMs depends on multiple factors such as network, device, etc., we only considered open-source LLMs in our %fairly
controlled computing environment. Tables \ref{tab_overview_b}-\ref{tab_overview_c} in our empirical study presented the cost analysis of four multi-agent frameworks on HumanEval. In this experiment, we analyzed their cost on MBPP dataset as shown in Tables \ref{tab_mbpp_tokens}-\ref{tab_mbpp_time}.

\begin{table*}
% \scriptsize
% \centering
% \caption{The resource consumption performance of AgentCoder, MapCoder, INTERVENOR and Self-Collaboration on MBPP.}
% \begin{subtable}[t]{\textwidth}
    \centering
    \scriptsize
    \caption{The average inference token consumption of AgentCoder, MapCoder, INTERVENOR and Self-Collaboration on MBPP.}
    \begin{tabular}{lcccccc}
        \toprule
        \textbf{LLMs} & \textbf{Direct} & \textbf{AgentCoder} & \textbf{MapCoder} & \textbf{INTERVENOR} & \textbf{Self-Collaboration} \\
        \midrule
        CodeLlama-Python-7B & 00.81K & 12.18K (\textcolor{cgreen}{$\uparrow$14.06$\times$}) & 32.46K (\textcolor{cgreen}{$\uparrow$39.16$\times$})& 14.09K (\textcolor{cgreen}{$\uparrow$16.43$\times$}) & 24.52K (\textcolor{cgreen}{$\uparrow$29.33$\times$}) \\
        CodeLlama-Python-13B & 00.76K & 12.40K (\textcolor{cgreen}{$\uparrow$15.37$\times$}) & 28.31K (\textcolor{cgreen}{$\uparrow$36.37$\times$})& 12.06K (\textcolor{cgreen}{$\uparrow$14.91$\times$}) & 18.88K (\textcolor{cgreen}{$\uparrow$23.91$\times$}) \\
        CodeLlama-Python-34B & 01.06K & 10.48K (\textcolor{cgreen}{$\uparrow$08.90$\times$}) & 24.74K (\textcolor{cgreen}{$\uparrow$22.37$\times$})& 09.77K (\textcolor{cgreen}{$\uparrow$08.22$\times$}) & 16.86K (\textcolor{cgreen}{$\uparrow$14.92$\times$}) \\
        DeepSeek-Coder-1.3B & 01.11K & 14.49K (\textcolor{cgreen}{$\uparrow$12.02$\times$}) & 33.06K (\textcolor{cgreen}{$\uparrow$28.70$\times$})& 15.25K (\textcolor{cgreen}{$\uparrow$12.70$\times$}) & 18.45K (\textcolor{cgreen}{$\uparrow$15.58$\times$}) \\
        DeepSeek-Coder-6.7B & 01.16K & 08.79K (\textcolor{cgreen}{$\uparrow$06.56$\times$}) & 37.78K (\textcolor{cgreen}{$\uparrow$31.50$\times$})& 09.50K (\textcolor{cgreen}{$\uparrow$07.18$\times$}) & 16.73K (\textcolor{cgreen}{$\uparrow$13.39$\times$}) \\
        DeepSeek-Coder-33B & 01.17K & 10.29K (\textcolor{cgreen}{$\uparrow$07.83$\times$}) & 30.95K (\textcolor{cgreen}{$\uparrow$25.56$\times$})& 11.46K (\textcolor{cgreen}{$\uparrow$08.84$\times$}) & 18.15K (\textcolor{cgreen}{$\uparrow$14.57$\times$}) \\
        \midrule
        $\Delta$ Average & - & \textcolor{cgreen}{$\uparrow$10.79$\times$} & \textcolor{cgreen}{$\uparrow$30.61$\times$} & \textcolor{cgreen}{$\uparrow$11.38$\times$} & \textcolor{cgreen}{$\uparrow$18.62$\times$} \\
        \bottomrule
    \end{tabular}
\label{tab_mbpp_tokens}
\end{table*}

\begin{table*}
    % \vspace{0.5cm}
    \scriptsize
    \centering
    \caption{The average inference time of AgentCoder, MapCoder, INTERVENOR and Self-Collaboration on MBPP.}
    \begin{tabular}{lccccc}
        \toprule
        \textbf{LLMs} & \textbf{Direct} & \textbf{AgentCoder} & \textbf{MapCoder} & \textbf{INTERVENOR} & \textbf{Self-Collaboration} \\
        \midrule
        CodeLlama-Python-7B  & 15.12s & 123.4s (\textcolor{cgreen}{$\uparrow$07.16$\times$}) & 319.5s (\textcolor{cgreen}{$\uparrow$20.14$\times$}) & 170.7s (\textcolor{cgreen}{$\uparrow$10.29$\times$}) & 226.9s (\textcolor{cgreen}{$\uparrow$14.01$\times$}) \\
        CodeLlama-Python-13B & 21.85s & 306.8s (\textcolor{cgreen}{$\uparrow$13.04$\times$}) & 515.0s (\textcolor{cgreen}{$\uparrow$22.57$\times$}) & 212.3s (\textcolor{cgreen}{$\uparrow$08.72$\times$}) & 345.2s (\textcolor{cgreen}{$\uparrow$14.80$\times$}) \\
        CodeLlama-Python-34B & 70.53s & 393.7s (\textcolor{cgreen}{$\uparrow$04.58$\times$}) & 916.1s (\textcolor{cgreen}{$\uparrow$11.99$\times$}) & 367.2s (\textcolor{cgreen}{$\uparrow$04.21$\times$}) & 611.8s (\textcolor{cgreen}{$\uparrow$07.67$\times$}) \\
        DeepSeek-Coder-1.3B  & 12.06s & 90.88s (\textcolor{cgreen}{$\uparrow$06.54$\times$}) & 182.1s (\textcolor{cgreen}{$\uparrow$14.10$\times$}) & 111.3s (\textcolor{cgreen}{$\uparrow$08.23$\times$}) & 97.33s (\textcolor{cgreen}{$\uparrow$07.07$\times$}) \\
        DeepSeek-Coder-6.7B  & 20.90s & 88.83s (\textcolor{cgreen}{$\uparrow$03.25$\times$}) & 359.8s (\textcolor{cgreen}{$\uparrow$16.22$\times$}) & 112.8s (\textcolor{cgreen}{$\uparrow$04.40$\times$}) & 194.6s (\textcolor{cgreen}{$\uparrow$08.31$\times$}) \\
        DeepSeek-Coder-33B   & 86.59s & 392.4s (\textcolor{cgreen}{$\uparrow$03.53$\times$}) & 1217.2s (\textcolor{cgreen}{$\uparrow$13.06$\times$}) & 541.5s (\textcolor{cgreen}{$\uparrow$05.25$\times$}) & 706.0s (\textcolor{cgreen}{$\uparrow$07.15$\times$}) \\
        \midrule
        $\Delta$ Average & - & \textcolor{cgreen}{$\uparrow$06.35$\times$} & \textcolor{cgreen}{$\uparrow$16.35$\times$} & \textcolor{cgreen}{$\uparrow$06.85$\times$} & \textcolor{cgreen}{$\uparrow$09.84$\times$} \\
        \bottomrule
    \end{tabular}
\label{tab_mbpp_time}
% \end{subtable}
% \label{tab_mbpp_resource}
\end{table*}

\vspace{5pt}\noindent\textbf{In terms of token consumption}, Table \ref{tab_AdaCoder_token} reveals that, compared to the foundation LLMs, AdaCoder shows a token increase of 2.00 and 2.42 times on HumanEval and MBPP, respectively, on average. The average increase of two datasets is 2.21 times. In contrast, MapCoder, the most effective baseline, leads to 23.02 and 30.61 times token increase on HumanEval and MBPP, respectively, on average. This results in an increase of 26.81 times among two datasets. Thus, MapCoder costs 12.13 times more tokens than AdaCoder for these two datasets. 

\vspace{5pt}\noindent\textbf{Regarding inference time}, Table \ref{tab_AdaCoder_time} shows that, compared to not using any multi-agent framework (i.e., Direct), using AdaCoder increases inference time by only 0.68 and 1.34 times (including the running time of the Code Evaluator and the Debug Specialist) on HumanEval and MBPP, respectively, with an average increase of 1.01 times. Meanwhile, MapCoder leads to an average increase of 15.72 and 16.35 times of inference time on HumanEval and MBPP, respectively, totaling an average of 16.04 times. Therefore, MapCoder requires 15.88X longer inference time than AdaCoder. These results demonstrate AdaCoder's low token consumption and inference time for code generation, compared with the best baseline. 

\begin{table}
    \centering
    % \scriptsize
    \caption{The average inference token consumption of AdaCoder on HumanEval and MBPP.}
    \fontsize{6}{7}\selectfont
    \begin{tabular}{l|c|c|c|c}
        \toprule
        \multirow{2}{*}{\textbf{LLMs}} & \multicolumn{2}{c|}{\textbf{Direct}} & \multicolumn{2}{|c}{\textbf{AdaCoder}} \\ 
        \cmidrule{2-5} & HumanEval & MBPP & HumanEval & MBPP \\
        \midrule
        CodeLlama-Python-7B  & 00.91K & 00.81K & 04.79K (\textcolor{cgreen}{$\uparrow$04.26$\times$}) & 03.29K (\textcolor{cgreen}{$\uparrow$03.07$\times$}) \\
        CodeLlama-Python-13B & 00.92K & 00.76K & 04.01K (\textcolor{cgreen}{$\uparrow$03.36$\times$}) & 03.10K (\textcolor{cgreen}{$\uparrow$03.09$\times$}) \\
        CodeLlama-Python-34B & 00.95K & 01.06K & 03.46K (\textcolor{cgreen}{$\uparrow$02.64$\times$}) & 03.06K (\textcolor{cgreen}{$\uparrow$01.89$\times$}) \\
        DeepSeek-Coder-1.3B  & 01.13K & 01.11K & 03.86K (\textcolor{cgreen}{$\uparrow$02.42$\times$}) & 03.81K (\textcolor{cgreen}{$\uparrow$02.42$\times$}) \\
        DeepSeek-Coder-6.7B  & 01.16K & 01.16K & 02.87K (\textcolor{cgreen}{$\uparrow$01.47$\times$}) & 03.23K (\textcolor{cgreen}{$\uparrow$01.78$\times$}) \\
        DeepSeek-Coder-33B   & 01.18K & 01.17K & 02.92K (\textcolor{cgreen}{$\uparrow$01.47$\times$}) & 03.20K (\textcolor{cgreen}{$\uparrow$01.75$\times$}) \\
        GPT-3.5-turbo        & 00.40K & 00.29K & 00.83K (\textcolor{cgreen}{$\uparrow$01.07$\times$}) & 01.08K (\textcolor{cgreen}{$\uparrow$02.72$\times$}) \\
        GPT-4                & 00.41K & 00.50K & 01.05K (\textcolor{cgreen}{$\uparrow$01.56$\times$}) & 01.76K (\textcolor{cgreen}{$\uparrow$02.54$\times$}) \\
        GPT-4-turbo          & 00.46K & 00.46K & 00.89K (\textcolor{cgreen}{$\uparrow$00.93$\times$}) & 01.57K (\textcolor{cgreen}{$\uparrow$02.42$\times$}) \\
        GPT-4o               & 00.37K & 00.41K & 00.66K (\textcolor{cgreen}{$\uparrow$00.78$\times$}) & 01.43K (\textcolor{cgreen}{$\uparrow$02.48$\times$}) \\
        \midrule
        $\Delta$ Average     & - & - & \textcolor{cgreen}{$\uparrow$02.00$\times$} & \textcolor{cgreen}{$\uparrow$02.42$\times$} \\
        \bottomrule
    \end{tabular}
\label{tab_AdaCoder_token}
\end{table}

\begin{table}
    \centering
    % \scriptsize
    % \vspace{0.5cm}
    \fontsize{6}{7}\selectfont
    \caption{The average inference time of AdaCoder on HumanEval and MBPP.}
    \begin{tabular}{l|c|c|c|c}
        \toprule
        \multirow{2}{*}{\textbf{LLMs}} & \multicolumn{2}{c|}{\textbf{Direct}} & \multicolumn{2}{|c}{\textbf{AdaCoder}} \\ 
        \cmidrule{2-5} & HumanEval & MBPP & HumanEval & MBPP \\
        \midrule
        CodeLlama-Python-7B  & 27.33s & 15.12s & 48.61s (\textcolor{cgreen}{$\uparrow$00.78$\times$}) & 32.88s (\textcolor{cgreen}{$\uparrow$01.18$\times$}) \\
        CodeLlama-Python-13B & 41.42s & 21.85s & 70.24s (\textcolor{cgreen}{$\uparrow$00.70$\times$}) & 69.80s (\textcolor{cgreen}{$\uparrow$02.20$\times$}) \\
        CodeLlama-Python-34B & 76.49s & 70.53s & 145.9s (\textcolor{cgreen}{$\uparrow$00.91$\times$}) & 144.7s (\textcolor{cgreen}{$\uparrow$01.05$\times$}) \\
        DeepSeek-Coder-1.3B  & 22.52s & 12.06s & 28.55s (\textcolor{cgreen}{$\uparrow$00.27$\times$}) & 30.38s (\textcolor{cgreen}{$\uparrow$01.52$\times$}) \\
        DeepSeek-Coder-6.7B  & 20.65s & 20.90s & 35.55s (\textcolor{cgreen}{$\uparrow$00.72$\times$}) & 43.29s (\textcolor{cgreen}{$\uparrow$01.07$\times$}) \\
        DeepSeek-Coder-33B   & 88.27s & 86.59s & 149.2s (\textcolor{cgreen}{$\uparrow$00.69$\times$}) & 176.4s (\textcolor{cgreen}{$\uparrow$01.04$\times$}) \\
        \midrule
        $\Delta$ Average     & - & - & \textcolor{cgreen}{$\uparrow$00.68$\times$} & \textcolor{cgreen}{$\uparrow$01.34$\times$} \\
        \bottomrule
    \end{tabular}
\label{tab_AdaCoder_time}
\end{table}

\vspace{0.3cm}
\begin{mdframed}[nobreak=true]
\textbf{Answer to RQ3:} AdaCoder demonstrates the best performance with high generalizability, significantly outperforming the best baseline MapCoder by 27.69\% on ten LLMs on average; its computation cost is low with 12.13 times less tokens and 15.88 times shorter inference time than MapCoder, respectively.
% Applying AdaCoder to both open-source and closed-source LLMs can enhance their performance by an average of 50.00\%, surpassing existing multi-agent frameworks. Meanwhile, AdaCoder only increases the average token consumption and inference time by 2.21 times and 1.01 times, respectively, which are approximately one twelfth and one sixteenth of those required by the exsiting most effective framework, MapCoder.
\end{mdframed}

\subsection{Ablation Study of AdaCoder's Agents (RQ4)}

\begin{table*}
    \scriptsize
    \centering
    \caption{Ablation study results of AdaCoder with different agents removed, applied to ten selected LLMs. ``w/o Prompt" indicates the removal of the Prompt Engineer, while ``w/o Debug" indicates the removal of the Debug Specialist.} 
    \begin{tabular}{lcccc}
        \toprule
        \textbf{LLMs} & \textbf{AdaCoder} & \textbf{w/o Prompt} & \textbf{w/o Debug} & \textbf{w/o Prompt \& Debug} \\
        \midrule
        CodeLlama-Python-7B & 63.41 & 48.78 (\textcolor{red}{$\downarrow$23.07\%}) & 58.54 (\textcolor{red}{$\downarrow$07.68\%}) & 41.46 (\textcolor{red}{$\downarrow$34.62\%}) \\
        CodeLlama-Python-13B  & 71.95 & 56.71 (\textcolor{red}{$\downarrow$21.18\%}) & 64.63 (\textcolor{red}{$\downarrow$10.17\%}) & 53.66 (\textcolor{red}{$\downarrow$25.42\%}) \\
        CodeLlama-Python-34B & 81.10 & 64.63 (\textcolor{red}{$\downarrow$20.31\%}) & 71.95 (\textcolor{red}{$\downarrow$11.28\%}) & 57.93 (\textcolor{red}{$\downarrow$28.57\%}) \\
        DeepSeek-Coder-1.3B & 76.83 & 62.20 (\textcolor{red}{$\downarrow$19.04\%}) & 70.12 (\textcolor{red}{$\downarrow$08.73\%}) & 54.27 (\textcolor{red}{$\downarrow$29.36\%}) \\
        DeepSeek-Coder-6.7B & 85.98 & 72.56 (\textcolor{red}{$\downarrow$15.61\%}) & 79.27 (\textcolor{red}{$\downarrow$07.80\%}) & 57.93 (\textcolor{red}{$\downarrow$32.62\%}) \\
        DeepSeek-Coder-33B & 90.85 & 74.39 (\textcolor{red}{$\downarrow$18.12\%}) & 70.73 (\textcolor{red}{$\downarrow$22.15\%}) & 61.59 (\textcolor{red}{$\downarrow$32.21\%}) \\
        GPT-3.5-turbo & 96.95 & 75.61 (\textcolor{red}{$\downarrow$22.01\%}) & 86.59 (\textcolor{red}{$\downarrow$10.69\%}) & 76.22 (\textcolor{red}{$\downarrow$21.38\%}) \\
        GPT-4 & 96.95 & 81.71 (\textcolor{red}{$\downarrow$15.72\%}) & 90.24 (\textcolor{red}{$\downarrow$06.92\%}) & 79.27 (\textcolor{red}{$\downarrow$18.24\%}) \\
        GPT-4-turbo & 98.17 & 91.46 (\textcolor{red}{$\downarrow$06.84\%}) & 92.07 (\textcolor{red}{$\downarrow$06.21\%}) & 90.24 (\textcolor{red}{$\downarrow$08.08\%}) \\
        GPT-4o & 98.17 & 91.46 (\textcolor{red}{$\downarrow$06.84\%}) & 93.90 (\textcolor{red}{$\downarrow$04.35\%}) & 89.02 (\textcolor{red}{$\downarrow$09.32\%}) \\
        \midrule
        $\Delta$ Average & - & \textcolor{red}{$\downarrow$16.87\%} & \textcolor{red}{$\downarrow$09.60\%} & \textcolor{red}{$\downarrow$23.98\%} \\
        \bottomrule
    \end{tabular}
\label{tab_ablation}
\end{table*}

% The experimental results are presented in Table \ref{tab_ablation}.

Table \ref{tab_ablation} shows that in all three scenarios, compared to the complete AdaCoder, the code generation capabilities of all LLMs decreased to varying degrees. Specifically, removing the Prompt Engineer resulted in a decrease in code generation capability ranging from 6.84\% to 23.07\% across the LLMs, with an average decrease of 16.87\%. CodeLlama-Python-7B experienced the largest decrease, while GPT-4-turbo and GPT-4o showed the smallest decrease. This phenomenon may be attributed to the latter two's inherently strong code generation capabilities, rendering the strategies designed by the Prompt Engineer less impactful. Removing the Debug Specialist led to a decrease in code generation capability ranging from 4.35\% to 22.15\%, with an average decrease of 9.60\%. DeepSeek-Coder-33B exhibited the largest decrease, while GPT-4o showed the smallest. This disparity might be due to GPT-4o being less prone to generating simple errors. Simultaneously removing both the Prompt Engineer and Debug Specialist resulted in a decrease in code generation capability ranging from 8.08\% to 34.62\%, with an average decrease of 23.98\%. CodeLlama-Python-7B experienced the largest decrease, while GPT-4-turbo showed the smallest. These results demonstrate the contributions of each component in AdaCoder.

\vspace{0.3cm}
\begin{mdframed}[nobreak=true]
\textbf{Answer to RQ4: } All agents individually contribute to the performance of AdaCoder, indicating their effectiveness and necessity to our design.
% Applying AdaCoder to both open-source and closed-source LLMs can enhance their performance by an average of 50.00\%, surpassing existing multi-agent frameworks. Meanwhile, AdaCoder only increases the average token consumption and inference time by 2.21 times and 1.01 times, respectively, which are approximately one twelfth and one sixteenth of those required by the exsiting most effective framework, MapCoder.
\end{mdframed}

\section{Threats to Validity}
We set the iteration count to five by default for AdaCoder's performance and demonstrate how the count can be adjusted to proportionally gain performance at the expense of time and token consumption in Figure \ref{fig_AdaCoder_t_value}. However, for other foundation LLMs and code generation tasks, this parameter may require some adjustment to achieve an optimal balance between pass@1 and resource consumption. Moreover, although our experiments are designed to mirror real-world scenarios, the specific datasets and foundation LLMs %employed 
may restrict the broader applicability of our findings. To mitigate this, we plan to validate our approach using more diverse datasets and in various environments in future research.

\section{Conclusion} \label{conclusion}
In this study, we first evaluate the generalizability of four state-of-the-art multi-agent frameworks %, such as MapCoder and Self-Collaboration, 
by applying them to six different LLMs from two families (i.e., CodeLlama-Python and DeepSeek-Coder). 
Our empirical findings on the HumanEval dataset reveal that their generalizability is unstable: MapCoder has the highest generalizability but with high inference cost. Its effectiveness can be attributed to its planning mechanism, i.e., Multi-Plan Coding, which guides LLMs in generating solutions through various plans. Subsequent analysis suggests that combining planning and non-planning mechanisms could achieve better performance and lower cost than using only the planning mechanism. In addition, iterative refinement process is both ineffective and costly.
% Subsequent analysis of the evaluation results and the underlying principles of each framework, conducted through multiple ablation experiments, led us to conclude that iteration refinement is both ineffective and costly. MapCoder excels due to its planning mechanism, ``Multi-Plan Coding'', despite a significant increase in token consumption and inference time. Additionally, its plan generation mechanism is unreliable, and the generated plans lack diversity. 

Motivated by these findings, we designed AdaCoder,  %\footnote{\url{https://github.com/YXingo/AdaCoder}},
an adaptive planning framework for multi-agent code generation. Evaluations demonstrate that AdaCoder achieves high generalizability compared to the best baseline MapCoder, surpassing it by 27.69\% in pass@1 while being applicable to LLMs of varying parameter scales, architectures, and performance levels. Furthermore, AdaCoder is 16 times faster and consumes 12 times fewer tokens than MapCoder. Additionally, ablation studies confirm the necessity of each agent in AdaCoder.
% Our evaluation of this framework demonstrated significant performance improvements for both the open-source (57.56\%) and closed-source (38.66\%) LLMs, surpassing existing best multi-agent frameworks MapCoder by 27.69\%.
% Meanwhile, the average token consumption and inference time of AdaCoder are only one twelfth and one sixteenth of those required by the existing most effective framework, MapCoder. 
% Meanwhile, AdaCoder also works efficiently because MapCoder consumes 12.13 times more tokens and requires 15.88 times longer inference time than AdaCoder, respectively.
% \lc{please modify this section correspondingly.}

%\hy{can add a URL to data and code}

Our source code and experimental data are available at \url{https://github.com/YXingo/AdaCoder}. 

% \bibliographystyle{ACM-Reference-Format}
% \bibliography{reference}

% \appendix

% \input{appendix}

% or


% \appendices
% \input{appendix}

% % you can choose not to have a title for an appendix
% % if you want by leaving the argument blank
% \section{}

% \section*{Acknowledgment}


\ifCLASSOPTIONcaptionsoff
  \newpage
\fi



%\bibliographystyle{IEEEtran}
\bibliographystyle{IEEEtran}
\bibliography{reference}

\newpage

\appendices
\section{Refinement Type Example} \label{appendix_refinement_type}

\begin{figure}[H]
    \centering
    \includegraphics[width=1\linewidth]{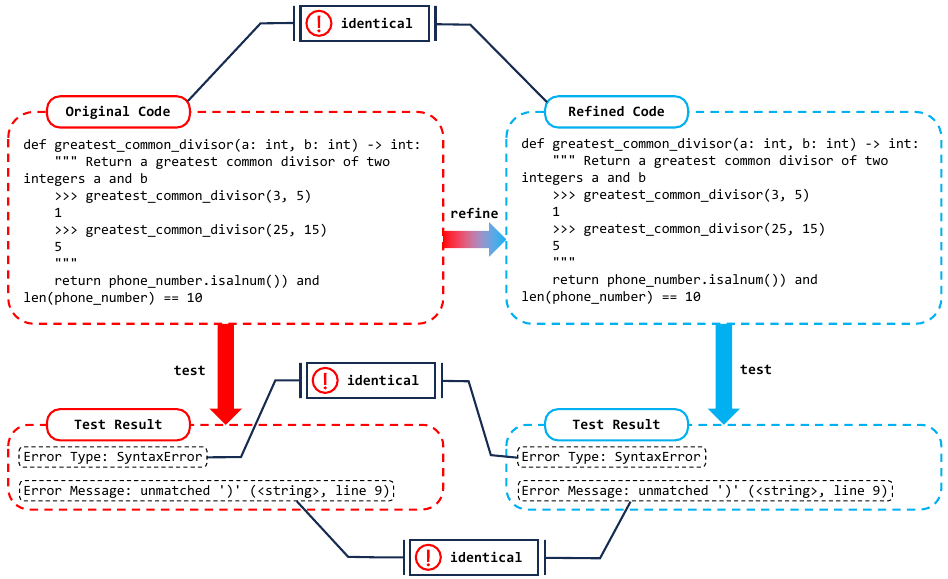}
    \caption{Example of the ``Code Invariance" situation in Table \ref{tab_refine}}
    \label{fig_code_invariance}
\end{figure}

\begin{figure}[H]
    \centering
    \includegraphics[width=1\linewidth]{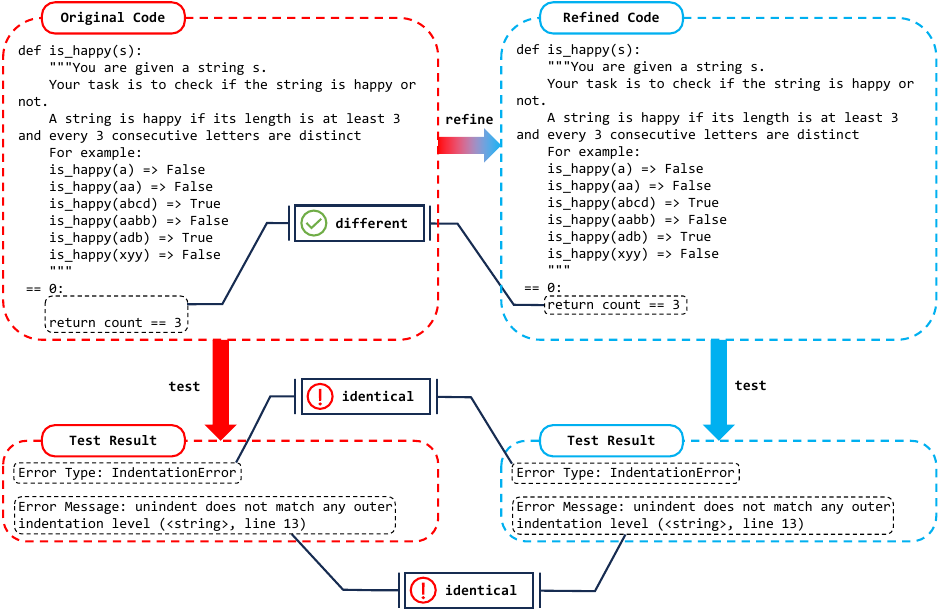}
    \caption{Example of the ``Error Message Persistence" situation in Table \ref{tab_refine}}
    \label{fig_error_message_persistence}
\end{figure}

\begin{figure}[H]
    \centering
    \includegraphics[width=1\linewidth]{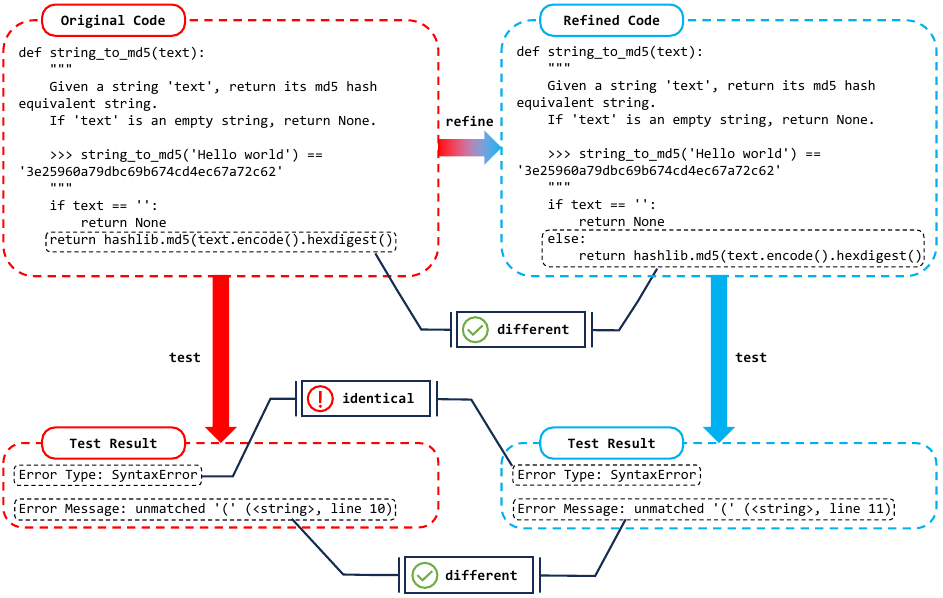}
    \caption{Example of the ``Error Type Consistency" situation in Table \ref{tab_refine}}
    \label{fig_error_type_consistency}
\end{figure}

\begin{figure}[H]
    \centering
    \includegraphics[width=1\linewidth]{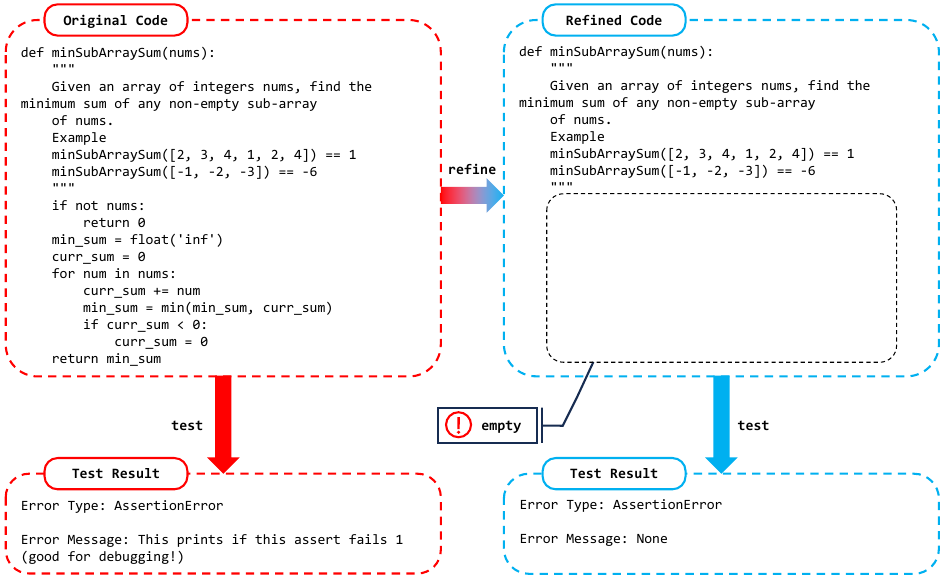}
    \caption{Example of the ``Function Emptying" situation in Table \ref{tab_refine}}
    \label{fig_function_emptying}
\end{figure}

\begin{figure}[H]
    \centering
    \includegraphics[width=1\linewidth]{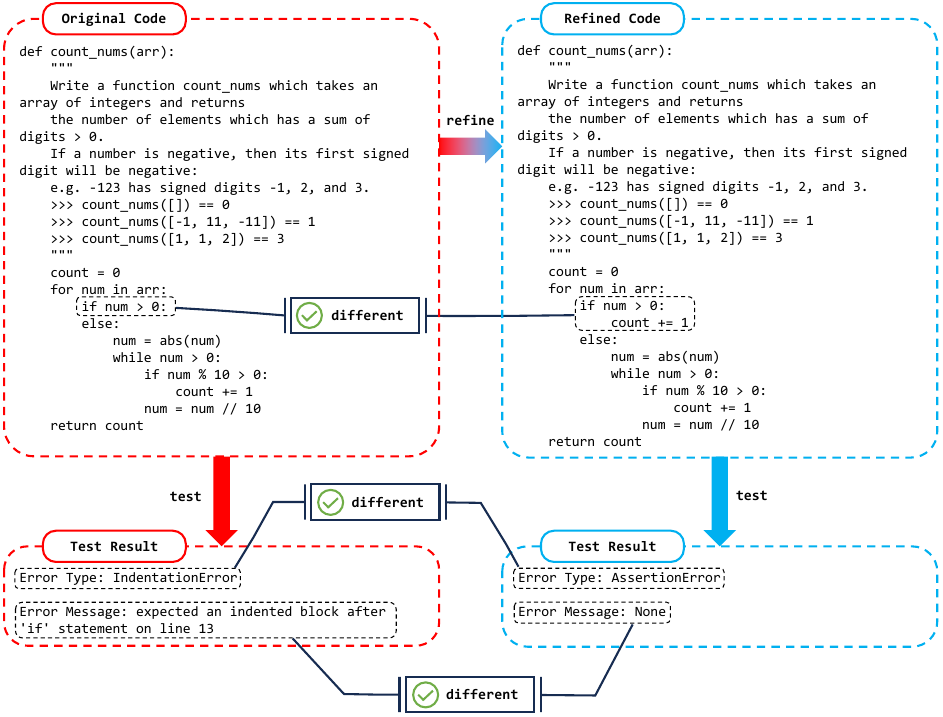}
    \caption{Example of the ``Miscellaneous Refinement" situation in Table \ref{tab_refine}}
    \label{fig_miscellaneous_refinement}
\end{figure}

\section{Detailed Prompting of AdaCoder} \label{appendix_prompt}

\begin{figure}[H]
    \centering
    \includegraphics[width=1\linewidth]{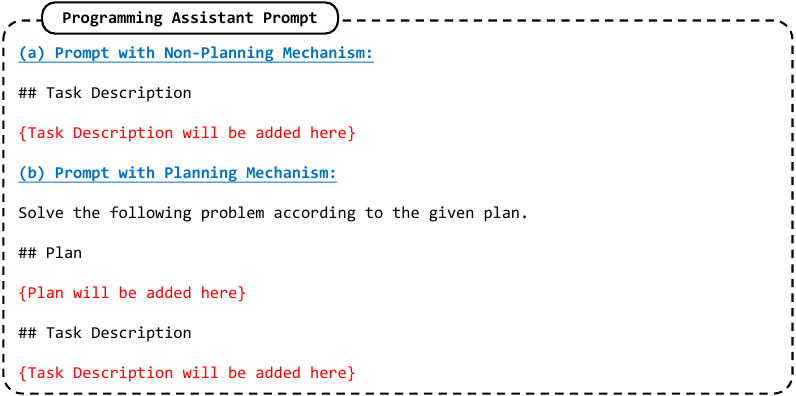}
    \caption{The prompt of the Programming Assistant.}
    \label{fig_PA_prompt}
\end{figure}

\begin{figure}[H]
    \centering
    \includegraphics[width=1\linewidth]{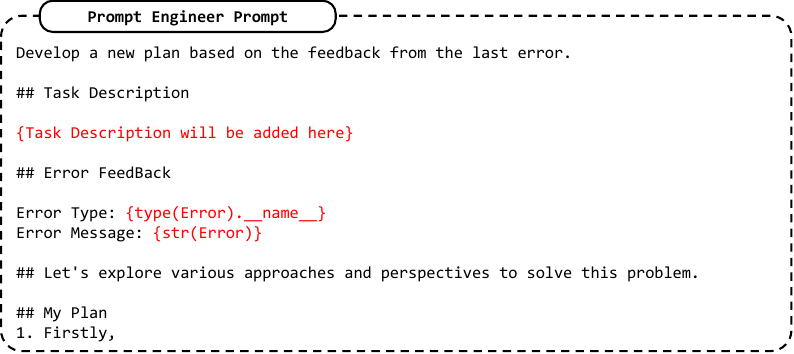}
    \caption{The prompt of the Prompt Engineer.}
    \label{fig_PE_prompt}
\end{figure}

\section{Pseudocode of AdaCoder and its two script agents}

\begin{algorithm}
\scriptsize
\caption{AdaCoder Workflow}
\label{algo_workflow}
\begin{algorithmic}[1]
\Require Task Description $T$, Maximum Iterations $t$
\Ensure Debugged Code $C_{\text{debugged}}$
\Procedure{AdaCoder}{$T$, $t$}
    \State Initialize $Plan \gets \text{None}$
    \For{$i \gets 0$ to $t$}
        \If{$Plan = \text{None}$}
            \State $C \gets \Call{ProgrammingAssistant}{T}$
        \Else
            \State $C \gets \Call{ProgrammingAssistant}{T, Plan}$
        \EndIf
        
        \State $EvalResult1 \gets \Call{CodeEvaluator}{T, C}$
        \If{$EvalResult1.\text{Status} = \text{Success}$}
            \State \Return $C$
        \EndIf
        
        \State $C_{\text{debugged}} \gets \Call{DebugSpecialist}{C, EvalResult1.\text{Info}}$
        
        \State $EvalResult2 \gets \Call{CodeEvaluator}{T, C_{\text{debugged}}}$
        \If{$EvalResult2.\text{Status} = \text{Success}$}
            \State \Return $C_{\text{debugged}}$
        \EndIf
        
        \State $Plan \gets \Call{PromptEngineer}{T, EvalResult2.\text{Info}}$
    \EndFor
    \State \Return $C_{\text{debugged}}$
\EndProcedure
\end{algorithmic}
\end{algorithm}

\begin{algorithm}
\scriptsize
\caption{Code Evaluator Workflow}
\label{alg_code_evaluator}
\begin{algorithmic}[1]
\Require Task Description $T$, Code Solution $C$
\Ensure Test Result $R$, Error Information $E$
\Procedure{CodeEvaluator}{$T$, $C$}
    \State Initialize $R \gets \text{Pass}$, $E \gets \text{None}$
    \State Retrieve $TestCases \gets \text{BenchmarkDataset}[T.\text{id}][\text{SampleTest}]$
    
    \State Embed $C$ in a \texttt{try-except} block for compilation
    \State \texttt{try:}
    \State \quad \Call{Compile}{$C$}
    \State \texttt{except CompilationError as err:}
    \State \quad $R \gets \text{Fail}$
    \State \quad $E \gets \text{err.message}$
    \State \quad \Return $\{R, E\}$
    
    \State ConcatenatedCode $\gets C$ + $TestCases$
    \State Embed $ConcatenatedCode$ in a \texttt{try-except} block for execution
    \State \texttt{try:}
    \State \quad \Call{Execute}{$ConcatenatedCode$}
    \State \texttt{except RuntimeError as err:}
    \State \quad $R \gets \text{Fail}$
    \State \quad $E \gets \text{err.message}$
    
    \State \Return $\{R, E\}$
\EndProcedure
\end{algorithmic}
\end{algorithm}

\begin{algorithm}
\scriptsize
\caption{Debug Specialist Workflow}
\label{alg_debug}
\begin{algorithmic}[1]
\Require Code Solution $C$, Error Information $E$
\Ensure Debugged Code $C_{\text{debugged}}$
\Procedure{DebugSpecialist}{$C$, $E$}
    \State Initialize $E \gets \text{None}$
    
    \State \textbf{Step 1: Code Filtering}
    \State $Lines \gets \Call{SplitLines}{C}$
    \ForAll{$Line \in Lines$}
        \If{$Line \neq \text{Empty}$}
            \State $LeadingSpaces \gets \Call{CountLeadingSpaces}{Line}$
            \State $CorrectedSpaces \gets \left\lfloor LeadingSpaces/4 \right\rfloor \times 4$
            \State $Line \gets \Call{ReplaceLeadingWhitespace}{Line, CorrectedSpaces}$
            \State $Line \gets \Call{ConvertSpacesToTabs}{Line}$
        \EndIf
    \EndFor
    
    \For{$i \gets 0$ to $\text{Length}(Lines) - 2$}
        \If{$\Call{EndsWithColon}{Lines[i]}$}
            \State $CurrentTabs \gets \Call{CountLeadingTabs}{Lines[i]}$
            \State $NextTabs \gets \Call{CountLeadingTabs}{Lines[i+1]}$
            \If{$NextTabs \leq CurrentTabs$}
                \State $Lines[i+1] \gets \Call{AddIndent}{Lines[i+1], CurrentTabs + 1}$
            \EndIf
        \EndIf
    \EndFor
    
    \State $C \gets \Call{JoinLines}{Lines}$
    \State $C \gets \Call{ReplaceIndentWithTabs}{C}$
    \State $C \gets \Call{RemoveExtraneousBlocks}{C}$
    
    \State \textbf{Step 2: Code Truncation}
    \While{$\Call{Compile}{C}$ fails \textbf{and} $\Call{FunctionCount}{C} > 1$}
        \State $C \gets \Call{RemoveLastRow}{C}$
        % \State \Call{AttemptCompile}{$C$}
    \EndWhile
    
    \State \textbf{Step 3: Missing Module Injection}
    \If{$\Call{Type}{E} = NameError$}
        \State $MissingSymbol \gets \Call{ParseMissingSymbol}{E}$
        \If{$MissingSymbol \in \Call{NameDatabase}{}$}
            \State $C \gets \Call{PrependImport}{C, MissingSymbol}$
        \EndIf
    \EndIf
    
    \State $C_{\text{debugged}} \gets C$
    
    \State \Return $C_{\text{debugged}}$
\EndProcedure
\end{algorithmic}
\end{algorithm}

\end{document}